\documentclass{article}

\usepackage[margin=1in]{geometry} 
\usepackage{graphicx,color,tikz,adjustbox,subfig,nicefrac,url}
\usepackage{amsmath,amsthm,amssymb}

\usepackage{pgfplots}

\pgfplotsset{
	width=10cm,
	compat=1.17
	}
	
\usepgfplotslibrary{fillbetween}

\usetikzlibrary{calc,arrows,backgrounds}

\usepackage{cite}

\renewcommand{\vec}[1]{\mathbf{#1}}

\newcommand{\tp}{{\mbox{\tiny \sf T}}}
\newcommand{\chd}{\mathcal C}
\renewcommand{\Re}{\mathbb{R}}

\newcommand{\T}{\mathcal T}
\newcommand{\W}{\mathcal W}
\newcommand{\Nint}{\mathcal N_{\text{int}}}
\newcommand{\Nlf}{\mathcal N_{\text{leaf}}}
\newcommand{\N}{\mathcal N}
\newcommand{\Q}{\mathcal Q}
\newcommand{\R}{\mathcal R}
\newcommand{\z}{\vec{z}}

\theoremstyle{plain}
\newtheorem{theorem}{Theorem}

\theoremstyle{definition}
\newtheorem{definition}[theorem]{Definition}

\newtheorem{problem}[theorem]{Problem}

\title{A Linear Programming Approach for Resource-Aware Information-Theoretic Tree Abstractions}

\author{Daniel T. Larsson\thanks{{D. Guggenheim} School of Aerospace Engineering, Georgia Institute of Technology, Atlanta, GA 30332{-0150}, USA; daniel.larsson@gatech.edu}%
\and Dipankar Maity\thanks{{Department} of Electrical and Computer Engineering, The University of North Carolina at Charlotte, Charlotte, NC 28223-0001, USA; dmaity@uncc.edu}%
\and Panagiotis Tsiotras\thanks{{D. Guggenheim} School of Aerospace Engineering, {Institute} for Robotics and Intelligent Machines, Georgia Institute of Technology, Atlanta, GA 30332{-0150}, USA; tsiotras@gatech.edu}}

\date{}

\begin{document}
	\maketitle
	
\begin{abstract}
	\noindent
	In this chapter, an integer linear programming formulation for the problem of obtaining task-relevant, multi-resolution, environment abstractions for resource-constrained autonomous agents is presented.
	The formulation leverages concepts from information-theoretic signal compression, specifically, the information bottleneck (IB) method, to pose an abstraction problem as an optimal encoder search over the space of multi-resolution trees.
	The abstractions emerge in a task-relevant manner as a function of agent information-processing constraints.
	We detail our formulation, and show how hierarchical tree structures, signal encoders, and information-theoretic methods for signal compression can be unified under a common theme.
	A discussion delineating the benefits and drawbacks of our formulation is presented, as well as a detailed explanation how our approach can be interpreted within the context of generating abstractions for resource-constrained autonomous systems.
	It is shown that the resulting information-theoretic abstraction problem over the space of multi-resolution trees can be formulated as a integer linear programming (ILP) problem.
	We demonstrate the approach on a number of examples, and provide a discussion detailing the differences of the proposed framework compared to existing methods.
	Lastly, we consider a linear program relaxation of the ILP problem, thereby demonstrating that multi-resolution information-theoretic tree abstractions can be obtained by solving a convex program.
\end{abstract}

\section{Introduction}

The identification and preservation of task-relevant information has long been considered central to the development of intelligent systems~\cite{Ponsen2010,Zucker2003}.
However, the design of task-relevant abstractions in autonomous systems has been traditionally handled by system engineers, who provide specific expert knowledge of the problem domain by specifying features that are important for the specific task at hand~\cite{Zucker2003,Ponsen2010,Holte2003}.
The use of abstractions in autonomous systems is motivated by the need to systematically reduce the complexity of problem-solving, such as planning and decision-making, by removing details that are considered irrelevant for the specific task~\cite{Holte2003,Zucker2003}.
As a result, the abstraction process allows autonomous agents to simplify the problem by focusing on aspects of the domain deemed to be task-relevant, while removing (or rather, ignoring) those details that are irrelevant~\cite{Holte2003,Zucker2003}.
Despite the fact that the ability to form task-relevant abstractions is considered central to the development of intelligent systems, many existing frameworks that leverage the power of hierarchical abstractions rely heavily on user-provided rules to form reduced representations of the problem domain~\cite{Zucker2003}.

\begin{figure}[h]
	\centering
	\subfloat[]{\includegraphics[width=0.49\textwidth]{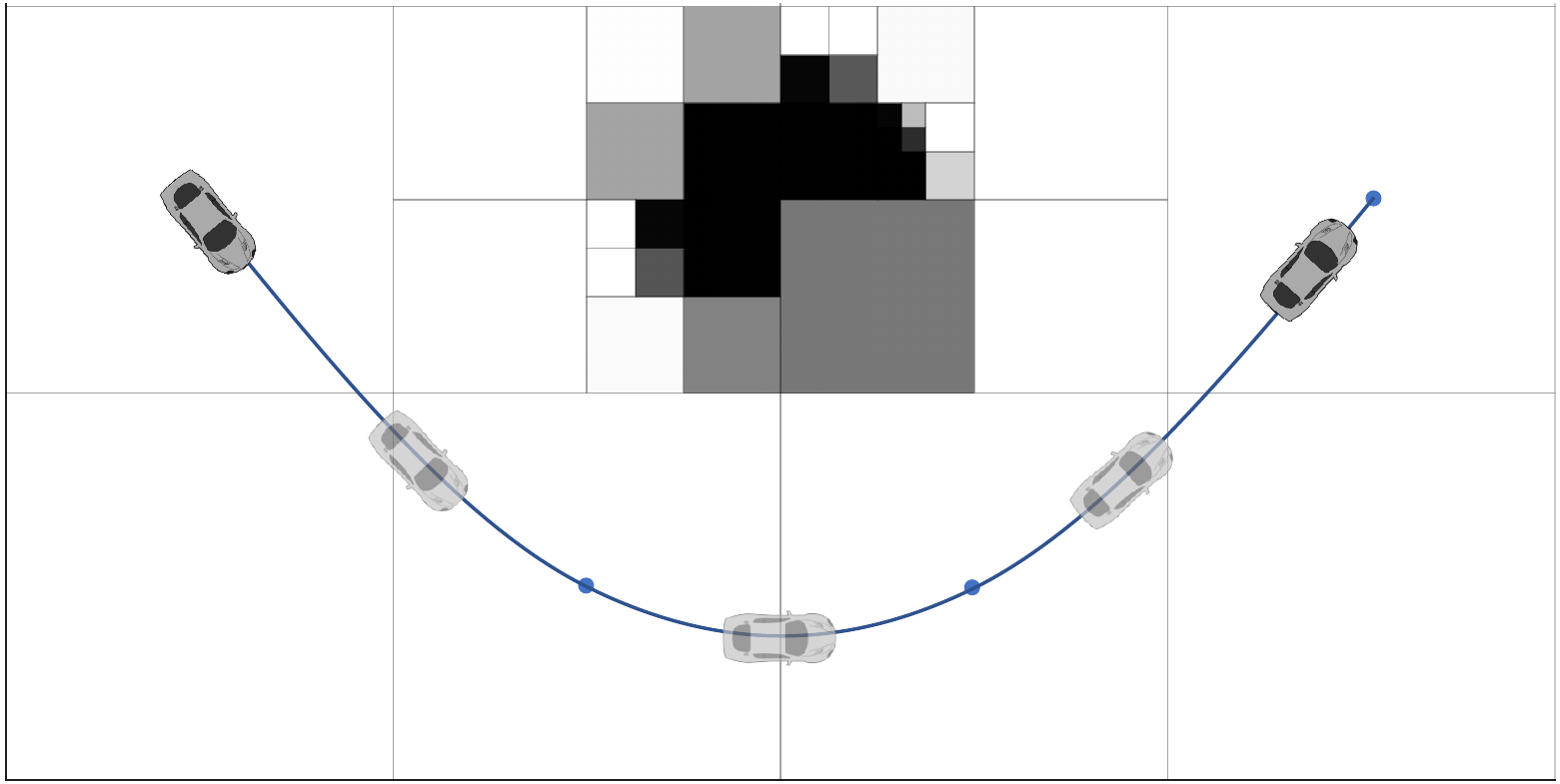}\label{fig:intro_ex_abs1}} \hfill
	\subfloat[]{\includegraphics[width=0.49\textwidth]{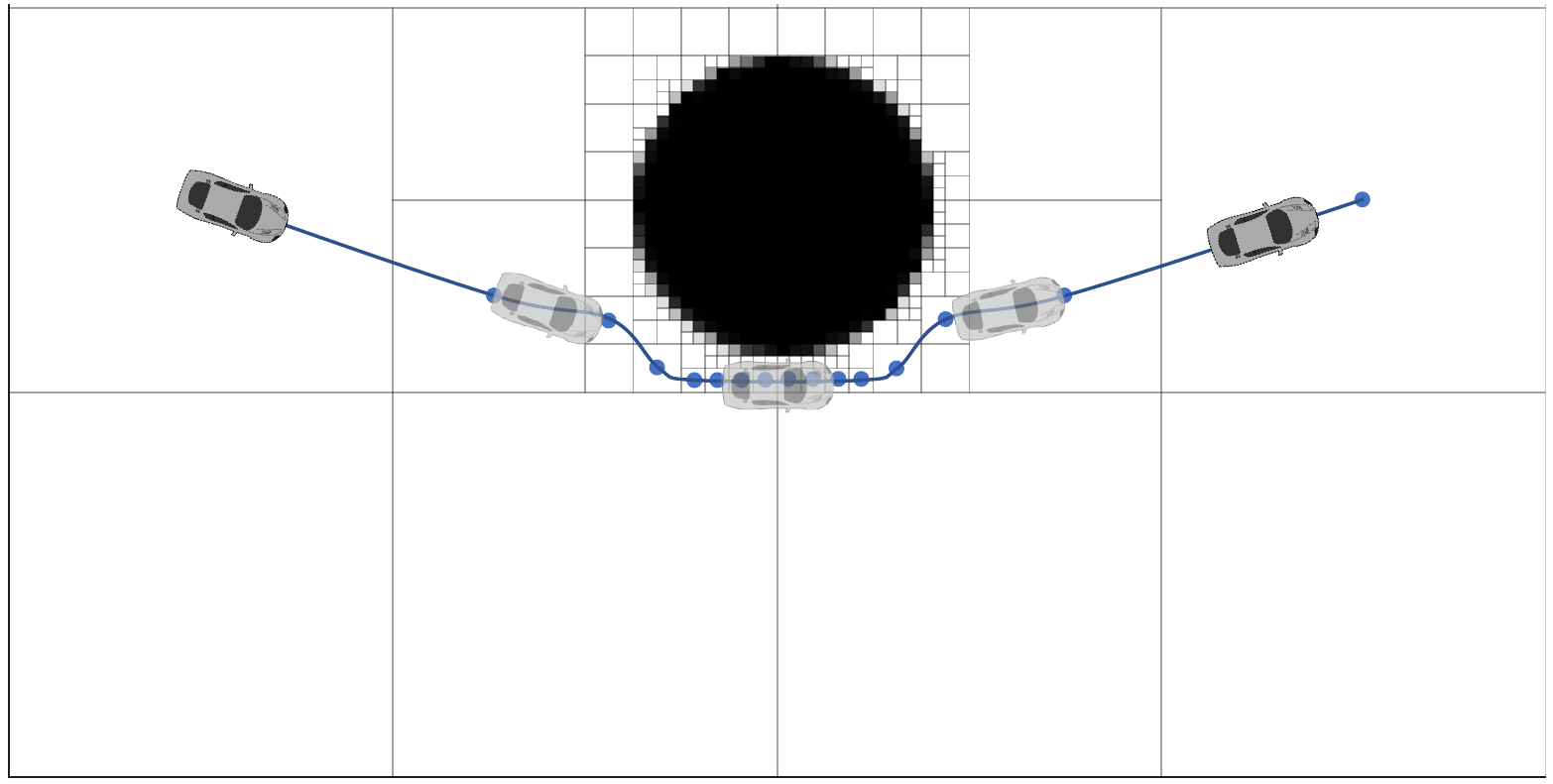}\label{fig:intro_ex_abs2}}
	\caption{Illustrative application of abstractions to the task of autonomous driving. 
	The darker cells represent  impassable obstacles in the environment.}
	\label{fig:intro_example_figure}
\end{figure}

Consider, for example, the task of planning a collision-free path in an autonomous driving scenario.
Assuming that a graph-based planning algorithm is employed to generate a path, such as A$^*$ or Dijkstra, an autonomous system must balance environment complexity (e.g., resolution) with the on-board resources available to execute the graph search in order to complete the task successfully in the presence of computational constraints.
To illustrate the trade-off between environment resolution and search complexity, consider the two environments shown in Fig.~\ref{fig:intro_example_figure}.
For each of the two abstractions shown in Fig.~\ref{fig:intro_example_figure}, one may construct a corresponding reduced graph whose vertices are the grid cell centers of the multi-resolution depiction and the corresponding edges encode cell connectivity.
As a result, executing graph search algorithms such as A$^*$ and Dijkstra on reduced graphs with fewer vertices, that is, those that correspond to lower-resolution representations, have better worst-case time complexity as compared with performing the graph-search on reduced graphs that are derived from higher-resolution depictions that contain a greater number of vertices~\cite{Larsson2020b}.
However, designing such abstractions is not trivial in that it requires the autonomous system to balance the model (environment) complexity with retaining as much information in regards to what is relevant to the task (in the example presented in Fig.~\ref{fig:intro_example_figure} this may be the occupancy probability of the finest-resolution grid cells).

Generating suitable compressed representations requires the formulation of a mathematical optimization problem that specifies which aspects of the problem are important (or, equivalently, relevant) and which can be neglected, or ignored, when forming a latent (e.g., compressed) representation of the problem.
Such a mathematical formulation requires the specification of two components.
The first component is a measure that captures the reduced model complexity, or rather, the degree of achieved compression compared to the original representation.
The second component is a measure that quantifies the quality of the abstract representation.
The terms driving compression and representation quality are in conflict with one another, in the sense that a higher-quality representation comes at the cost of a more complex abstract space, and vice-versa.
The design of abstractions is then formally posed as an optimization problem where the goal is to maximize compression subject to constraints on the reduced model quality.

The construction of the abstract, or compressed, representations of the environment is then subject to the definition of the measures that capture model complexity and quality.
By changing the measures that capture these two quantities, we arrive at different abstraction frameworks.
It should be noted that the specification of measures that capture model complexity and quality are often not trivial for a given task, but are vital to the construction of a good abstract space, as they implicitly define what details should be retained, and which can be removed, when forming a compressed representation.
To illustrate this trade-off, consider the framework of rate-distortion theory, which was developed by information theorists in order to encode signals for transmission over noisy communication channels~\cite{Cover2006}.
Rate-distortion theory formulates an optimization problem where the goal is to maximize compression subject to an upper bound on the expected distortion (i.e., bound the desired model quality).
Compressed representations that achieve lower values of expected distortion are considered to be of higher quality.
In its formulation, the rate-distortion framework measures the model complexity by utilizing Shannon's mutual information (MI) between the original and compressed signals, and quantifies the model quality through a non-negative distortion function.
The use of mutual information as a measure of the model complexity is rooted in its use in communication systems, where it quantifies the rate of codes required to describe a signal to a given level of detail~\cite{Cover2006}.
Crucially, however, rate-distortion theory does not specify the distortion measure, nor does it provide with any guidance  how one should be constructed, assuming instead that the function is provided by the system designer~\cite{Tishby1999}.
The selection of the distortion function is crucial, as it defines what details of the original signal should be retained in order to achieve low values of expected distortion (high model quality).
Specifying a distortion function for a given clustering task is difficult and non-intuitive, as it may not be clear from the problem formulation which features of the data should be preserved to accomplish a desired objective~\cite{Tishby1999}.

In order to combat some of the difficulties surrounding the specification of a distortion function, the work of~\cite{Tishby1999} introduced the Information-Bottleneck (IB) method.
In contrast to the approach employed by classical rate-distortion theory, the IB method does not rely on the specification of a distortion function, and instead introduces a third variable, referred to as the relevant random variable, that contains task-relevant information we wish to retain through the process of abstraction.
Consequently, the IB formulation replaces the distortion measure with the mutual information between the relevant variable and compressed representation.
Considering their similarities, it is unsurprising that the IB method has been informally viewed as a specific instance of the rate-distortion problem, where the distortion measure is taken to be that of the mutual information between the compressed and relevant random variables~\cite{GiladBachrach2003,Strouse2017,Tishby1999,Chechik2002}.
By virtue of its formulation, the IB problem naturally captures the notion of designing abstractions that are maximally retentive regarding task-specific relevant information.

In the autonomous systems community, abstractions have seen widespread use as a way to reduce the complexity of decision-making and path-planning problems~\cite{Tsiotras2007,Tsiotras2011,Hauer2019,Hauer2015,Cowlagi2012,Cowlagi2011,Cowlagi2010,Cowlagi2008}.
For example, in the works of~\cite{Tsiotras2007,Tsiotras2011,Hauer2019,Hauer2015,Cowlagi2012,Cowlagi2011,Cowlagi2010,Cowlagi2008}, the authors create coarse-grained, multi-resolution, representations of an environment for path-planning to reduce the computational complexity of executing graph-search algorithms.
The idea is that, by leveraging abstractions to create a reduced graph of the world containing fewer vertices than the full-resolution map, the execution time of off-the-shelf graph-search algorithms, such as A\(^*\) and Dijkstra, can be improved at the cost of sub-optimal paths.
In designing the abstractions, the works of~\cite{Tsiotras2007,Tsiotras2011,Hauer2019,Hauer2015,Cowlagi2012,Cowlagi2011,Cowlagi2010,Cowlagi2008} consider the relevant information as the region nearest the autonomous agent, as autonomous agents do not generally need to represent environment information that is physically distant from the vehicle with high resolution, as it is not crucial for short-range obstacle avoidance and sensing.
To strike a balance between (path) optimality and environment complexity the planning frameworks in~\cite{Tsiotras2007,Tsiotras2011,Hauer2019,Hauer2015,Cowlagi2012,Cowlagi2011,Cowlagi2010,Cowlagi2008} sequentially re-solve the planning problem as the agent traverses the world.
Crucially, the definition of information relevance employed by these works is somewhat arbitrary, and their reliance on the specification of tuning parameters that characterize the properties of the abstractions is a notable drawback.

Our goal in this chapter is to utilize the powerful approaches for signal compression to rigorously design task-relevant environment abstractions for autonomous systems by appealing to methods from information theory.
We aim to develop frameworks for the emergence of environment abstractions in autonomous systems that are task-driven and require minimal domain knowledge or interaction with the system designer.
To accomplish our goal, we propose to utilize concepts from information theory, specifically the IB principle, in order to formulate optimal encoder problems that give rise to task-relevant abstractions as a function of agent resource limitations.
The proposed method aims to afford agents the ability to design abstractions autonomously as a function of the task, while considering system-level resource constraints in a principled and rigorous manner.

\section{Notation}
Throughout this chapter, let the set of real numbers be denoted \(\Re\), and, for any strictly positive integer \(n\), let \(\Re^n\) denote the \(n\)-dimensional Euclidean space.
For any vector \(\vec{x} \in \Re^n\), \([\mathbf x]_i\) denotes the \(i^{\text{th}}\) element of \(\vec{x}\) for \(i \in \{1,\ldots,n\}\).
Furthermore, for any two vectors \(\vec{x},\vec{y} \in \Re^n\), the notation \(\vec{x} \leq \vec{y}\) is to be understood component-wise; that is, \([\vec{x}]_i \leq [\vec{y}]_i\) for all \(i \in \{1,\ldots,n\}\).
The expected value of the discrete random variable \(X\) distributed according to $p(x)$ is denoted as \(\mathbb E( X ) = \sum_x p(x) x\).
We assume throughout that logarithms have base \(e\).

\section{Preliminaries}

Our objective in this chapter is to develop a framework for the design of multi-resolution, task-relevant, tree abstractions of operating environments for resource-constrained autonomous systems.
To accomplish our goal, we first briefly review the fundamentals of information-theoretic signal compression by considering two notable frameworks: rate-distortion theory and the information bottleneck (IB) method; the latter will form the basis for the remainder of this chapter.
We then review hierarchical tree structures, and detail the connection between trees and signal encoders.
Facilitating a connection between hierarchical structures and encoders allows us to utilize  information-theoretic signal compression theory to formalize an abstraction problem over the space of multi-resolution trees.

\subsection{Information-Theoretic Signal Compression} \label{subsec:ITSignalCompression}

A number of frameworks have been developed by information theorists in order to compress signals for transmission across capacity-limited communication channels.
Two notable such frameworks are  the classical rate-distortion theory~\cite{Cover2006,Slonim2002}, and that of the information-bottleneck method~\cite{Tishby1999}.
Both of these approaches consider the design of optimal encoders by modeling signals probabilistically, employing concepts from probability theory to formulate rigorous optimization problems.
Therefore, in order to describe these frameworks in more detail, we require the introduction of a probability space \((\Omega, \mathcal F, \mathbb P)\), with  finite sample space \(\Omega\), \(\sigma\)-algebra \(\mathcal F\) and probability measure \(\mathbb P: \mathcal F \to [0,1]\).
Let \(X: \Omega \to \Re\) be a random variable defined on the sample space \(\Omega\) with distribution \(p(x) = \mathbb P(\{\omega \in \Omega : X(\omega) = x\}) \).
Similarly, we define the random variables \(T: \Omega \to \Re\) and \(Y: \Omega \to \Re\) with analogous definitions for the distributions \(p(t)\) and \(p(y)\).
We will give concrete meaning to the random variables \(X\), \(T\), and \(Y\) when discussing the compression frameworks below, but will regard them as general random variables for the time being.

Given a random variable \(X\) with associated distribution \(p(x)\), the \textit{(Shannon) entropy} of \(X\) is defined according to
\begin{equation} \label{eq:ShannonEntropy}
	H(X) = - \sum_{x} p(x) \log p(x).
\end{equation}
The Shannon entropy is non-negative and quantifies the uncertainty in the outcome of the random variable \(X\). 
More specifically, the Shannon entropy attains its maximum value when \(p(x)\) assigns equal probability mass to each outcome of \(X\), and where its minimal value of \(H(X) = 0\) is realized if \(p(x)\) assigns all probability mass to a single outcome (i.e., \(X\) is deterministic).
Notice that the Shannon entropy does not depended on the outcomes of the random variable \(X\), but rather only on its distribution \(p(x)\).
For this reason, we will at times abuse notation and write \(H(p)\) to mean \(H(p) = - \sum_x p(x) \log p(x)\).
While the entropy quantifies the uncertainty in \(X\), it may also be useful to have a means to measure the average uncertainty in \(X\) when given the outcome of another random variable, say \(T\).
To this end, the conditional entropy of the random variable \(X\) given \(T\) is defined as
\begin{equation} \label{eq:ShannonCondEntropy}
	H(X|T) = - \sum_{t,x} p(t,x) \log p(x|t).
\end{equation}
If the random variables \(X\) and \(T\) are independent, such that knowing the outcome of \(T\) does not influence our knowledge of \(X\), then \(H(X|T) = H(X)\).
Moreover, it can be shown that \(H(X|T) \leq H(X)\), representing the intuitive notion  that additional information cannot increase the uncertainty in a random variable~\cite{Cover2006}.

A related information-theoretic quantity is that of the Kullback-Liebler (KL) divergence, which quantifies the difference between two distributions \(p(x)\) and \(\nu(x)\).
Formally, the KL-divergence is defined as
\begin{equation} \label{eq:KLdivergence}
	\mathrm{D}_{\mathrm{KL}}(p(x),\nu(x)) = \sum_x p(x) \log \frac{p(x)}{\nu(x)}.
\end{equation}
The KL-divergence is, however, not a metric, as it is not symmetric and does not satisfy the triangle inequality~\cite{Cover2006,Slonim2002}.
It should be noted that despite these technical considerations, the KL-divergence is zero,  \(\mathrm{D}_{\mathrm{KL}}(p(x),\nu(x)) = 0\), if and only if \(p(x) = \nu(x)\), and is strictly positive otherwise.
The mutual information between two random variables \(X\) and \(T\) is defined in terms of the KL-divergence as
\begin{equation} \label{eq:mutualInfoDef}
	I(T;X) = \mathrm{D}_{\mathrm{KL}}(p(t,x), p(t) p(x)) = \sum_{t,x} p(t,x) \log \frac{p(t,x)}{p(t) p(x)}.
\end{equation}
From the definition of KL-divergence, we see that \(I(T;X) \geq 0\) and that \(I(T;X) = 0\) if and only if \(p(t,x) = p(t)p(x)\).
Consequently, if \(I(T;X) = 0\), it follows that the two random variables \(T\) and \(X\) are independent.
The mutual information $I(T;X)$ can expressed in terms of the Shannon entropy as
\begin{equation} \label{eq:mutualInfoEntropy}
	I(T;X) = H(T) - H(T|X) = H(X) - H(X|T).
\end{equation}
Relation \eqref{eq:mutualInfoEntropy} shows that the mutual information is the \textit{reduction in uncertainty} regarding the random variable \(T\) when \(X\) is known, and vice-versa~\cite{Cover2006}.
Consequently, if \(I(X;T) = 0\) then there is no reduction in uncertainty of \(T\) when given \(X\), corresponding to the case when \(T\) and \(X\) are independent; that is,  \(I(T;X) = 0\) means that \(T\) conveys no information regarding \(X\).
The above arguments can be equivalently interpreted by replacing \(X\) with \(T\), as the mutual information is symmetric, i.e., \(I(T;X) = I(X;T)\).
Furthermore, equation \eqref{eq:mutualInfoEntropy} along with the non-negativity of entropy furnish the inequality conditions \(0 \leq I(T;X) \leq H(T)\) and, similarly, \(0 \leq I(T;X) \leq H(X)\).
As we will see, the mutual information plays a central role in signal compression problems developed by information theorists.

The development of signal compression problems requires two components: (i) a measure that quantifies the degree of compression achieved by the reduced model, and (ii) a method to quantify the quality of the reduced model which captures how well the compressed representation represents the original model.
These two terms, the model complexity and quality of compression, are in direct conflict in the sense that a higher-quality model comes at the expense of a lower degree of achieved compression.
Forming high-quality abstractions is then a matter of identifying which information in a signal is relevant and should be preserved, and what is irrelevant and can be safely discarded.

To rigorously formalize these notions, we let the random variable \(X\) represent the original, uncompressed, signal and take the random variable \(T\) to represent the compressed representation of \(X\).
In addition, let \(\Omega_X = \{x\in\Re : X(\omega) = x, ~\omega \in \Omega\}\) and \(\Omega_T = \{t \in \Re : T(\omega) = t, ~\omega \in \Omega\}\) be the outcome spaces of \(X\) and \(T\), respectively.
Given the distribution \(p(x)\) of the original signal, the framework of rate-distortion theory formalizes the optimal encoder selection problem as
\begin{equation} \label{eq:RDTheoryProblem}
	\min_{p(t|x)} I(T;X),
\end{equation}
subject to the constraint
\begin{equation} \label{eq:RDTheoryProblemCons}
	\mathbb E ( d(T,X) ) \leq \tilde D,
\end{equation}
where \(d: \Omega_T \times \Omega_X \to [0,\infty)\) is a provided distortion function, which is assumed to take on small values when \(T\) is a good representation of \(X\), and \(\tilde D \geq 0\) is a given non-negative constant.
The problem~\eqref{eq:RDTheoryProblem} subject to~\eqref{eq:RDTheoryProblemCons} is over all normalized distributions (encoders) \(p(t|x)\) which map outcomes of \(X\) to outcomes of \(T\), and where the expectation in the constraint relation~\eqref{eq:RDTheoryProblemCons} is with respect to the joint distribution \(p(t,x) = p(t|x) p(x)\).
The constrained optimization problem \eqref{eq:RDTheoryProblem}-\eqref{eq:RDTheoryProblemCons} has Lagrangian
\begin{equation} \label{eq:RDTheoryLagrangian}
	\min_{p(t|x)} I(T;X) + \lambda ~\mathbb E (d(T,X) ),
\end{equation}
where \(\lambda \geq 0\) is a Lagrange multiplier whose value depends on \(\tilde D \).
For a given value of \(\lambda \geq 0\), the problem \eqref{eq:RDTheoryLagrangian} can be solved analytically, resulting in the pair of formal, self-consistent equations given by
\begin{align}
	p(t|x) &= \frac{p(t)}{\bar Z(x;\lambda)} \exp(-\lambda d(t,x)) \label{eq:optEncoderRDTheory}, \\
	p(t) &= \sum_x p(t|x) p(x), \label{eq:optMarginalRDTheory}
\end{align}
where \( \bar Z(x;\lambda) = \sum_{t} p(t) \exp(-\lambda d(t,x)) \)~\cite{Tishby1999,Cover2006,Slonim2002}.
The relations \eqref{eq:optEncoderRDTheory}-\eqref{eq:optMarginalRDTheory} are formal due to the dependence of \(p(t)\) on the conditional distribution \(p(t|x)\).
Despite the fact that the relations do not yield an explicit solution, a numerical approach can be employed, known as the Blahut-Arimoto (BA) algorithm, to obtain a global solution to the rate-distortion problem~\cite{Cover2006}.

In the rate-distortion framework, the degree of achieved compression is quantified by the mutual information \(I(T;X)\) between the compressed representation and original signal.
The use of \(I(T;X)\) as a measure of model complexity is rooted in communication theory, where \(I(T;X)\) is the rate, or rather, the average number of bits per message required to unambiguously specify an element in the codebook~\cite{Tishby1999}.
Consequently, lower rates imply that \(T\) is more compressed version of \(X\).
Moreover, observe that an encoder \(p(t|x)\) resulting in full compression \(I(T;X) = 0\) can always be obtained by removing all the details of a given signal \(X\).
An encoder \(p(t|x)\) achieving full compression renders \(T\) and \(X\) statistically independent, and incurs a high distortion cost.
Thus, by constraining the amount of allowable distortion, the framework must identify which aspects of the original signal are the most crucial to retain in order to maintain low distortion when designing the encoder \(p(t|x)\).

To more clearly understand how the distortion function conveys which aspects of the signal are relevant, we inspect the formal solution in equation~\eqref{eq:optEncoderRDTheory} for a given value of \(\lambda \geq 0\) and \(x \in \Omega_X\).
Here, we see that those outcomes \(t\in \Omega_T\) that result in lower distortion \(d(t,x)\) are assigned greater probability mass as compared to those outcomes \(t'\in\Omega_T\) that result in higher distortion \(d(t,x) < d(t',x)\).
Thus, the encoder \(p(t|x)\) will assign greater probability mass to those outcomes \(t \in \Omega_{T}\) which incur smaller values of \(d(t,x)\) for each \(x \in \Omega_X\) and \(\lambda \geq 0\).
It should be noted that the value of \(\lambda \geq 0\) is dependent on \(\tilde D\) in the constraint~\eqref{eq:RDTheoryProblemCons}~\cite{Cover2006}.
A notable drawback of the rate-distortion framework, however, lies in the need to specify the distortion function $d(t,x)$, which implicitly specifies which aspects of a signal are relevant, and should be preserved through the process of compression.
Furthermore, the specification of a distortion function for a given task may be difficult and non-intuitive~\cite{Tishby1999}.

In contrast, the information-bottleneck method~\cite{Tishby1999,GiladBachrach2003} formulates an optimal encoder problem that does not require the specification of a distortion function.
Instead, the IB problem aims to maximize the compression while explicitly constraining the amount of task-relevant information that is retained in the compressed representation.
To formalize these notions, we take the random variable \(X\) to represent the original signal and denote its compressed representation by \(T\).
In addition, the IB problem introduces a third, relevant random variable, which we will denote by \(Y\).
The random variable \(Y\) is assumed to be correlated with \(X\) and contain the information that we wish to preserve when forming the compressed representation \(T\).
The IB method assumes that the joint distribution \(p(x,y)\) is provided, and considers the problem
\begin{equation} \label{eq:IBproblemMinOpt}
	\min_{p(t|x)} I(T;X),
\end{equation}
subject to the constraint
\begin{equation} \label{eq:IBproblemMinOptCons}
	I(T;Y) \geq \hat D,
\end{equation}
where optimization is over normalized distributions \(p(t|x)\) and \(\hat D \geq 0\) is a given constant.
The joint distribution \(p(t,x,y)\) factors according to \(p(t,x,y) = p(t|x) p(x,y)\), which implies\footnote{As \(T\) is a compressed representation of \(X\), the IB problem searches for a mapping between outcomes of \(X\) and those of \(T\). The resulting mapping (encoder) is not a function of the relevant variable \(Y\), and so \(p(t|x,y) = p(t|x)\) or \(T \leftarrow X \leftarrow Y\). It can be shown that \(T \leftarrow X \leftarrow Y\) implies \(T \rightarrow X \rightarrow Y\) and thus \(T \leftrightarrow X \leftrightarrow Y\).} \(T \leftrightarrow X \leftrightarrow Y\) and, consequently, \(I(T;Y) \leq I(X;Y)\),  which follows from the data-processing inequality~\cite{Cover2006}.
The IB problem was first introduced in~\cite{Tishby1999}, where the authors considered the problem
\begin{equation} \label{eq:IBminProbLagrangian}
	\min_{p(t|x)} I(T;X) - \beta I(T;Y),
\end{equation}
where \(\beta \geq 0\) is a scalar whose value depends on \(\hat D\)~\cite{GiladBachrach2003}.
For a given value of \(\beta \geq 0\), the problem \eqref{eq:IBminProbLagrangian} can be solved analytically, resulting in the following formal relations that characterize the optimal solution
\begin{align}
	p(t|x) &= \frac{p(t)}{Z(x;\beta)} \exp (-\beta \mathrm{D}_{\mathrm{KL}}(p(y|x),p(y|t)) ), \label{eq:IBencoderSol}\\
	p(y|t) &= \sum_x p(y|x) p(x|t), \label{eq:IBcondDistSol} \\
	p(t) &= \sum_x p(t|x)p(x), \label{eq:IBmarginalSol}
\end{align}
where \( Z(x;\beta) = \sum_t p(t) \exp (-\beta \mathrm{D}_{\mathrm{KL}}(p(y|x),p(y|t)) \)~\cite{Tishby1999,Slonim2002}.
As in rate-distortion theory, the solution to the IB problem, characterized by relations \eqref{eq:IBencoderSol}-\eqref{eq:IBmarginalSol}, is formal as both the distributions \(p(t)\) and \(p(y|t)\) depend on the encoder \(p(t|x)\).
Moreover, comparing the relations~\eqref{eq:optEncoderRDTheory} and~\eqref{eq:IBencoderSol}, we see that the optimal solutions to the rate-distortion problem and IB method are very similar, and that the KL-divergence \(\mathrm{D}_{\mathrm{KL}}(p(y|x),p(y|t))\) appears in place of a distortion function \(d(t,x)\).
Recall, however, that in contrast to rate-distortion theory, the IB method does not rely on the specification of a distortion function, instead requiring only the specification of the statistical distribution \(p(x,y)\).
It is therefore important to note that the KL-divergence \(\mathrm{D}_{\mathrm{KL}}(p(y|x),p(y|t))\) in~\eqref{eq:IBencoderSol} \textit{emerges} as a distortion-like function from the analytical derivation of the solution to \eqref{eq:IBminProbLagrangian}, and is not specified or provided a priori~\cite{Tishby1999,Slonim2002}.
To conclude this section, we note that a solution to the IB problem~\eqref{eq:IBminProbLagrangian} may be found as a function of \(\beta\) by employing a numerical approach that utilizes the relations~\eqref{eq:IBencoderSol}-\eqref{eq:IBmarginalSol} to construct an iterative method similar to the BA algorithm in rate-distortion theory, although convergence of the IB algorithm is guaranteed only to local solutions~\cite{Tishby1999,Slonim2002}.

\subsection{Hierarchical Trees \& Trees as Deterministic Encoders} \label{subsec:treesAsEncoders}

Our goal in this chapter is to develop a framework for designing abstractions that endow autonomous systems the ability to trade environment complexity and model quality in a principled manner.
To accomplish our goal, in the previous section we provided an overview of information-theoretic signal compression approaches that design abstract representations of general signals with fidelity guarantees.
Importantly, the encoders that emerge as a result of the frameworks discussed in Section~\ref{subsec:ITSignalCompression} do not impose any structural constraints on the resulting solution \(p(t|x)\).
However, as we will see, hierarchical multi-resolution tree abstractions can be realized as encoders \(p(t|x)\) with very specific structure, and dealing with this additional structural constraint renders the resulting problem difficult to solve.
In order to make these notions concrete, we introduce in this section the formalism of hierarchical tree structures and discuss their connection to encoder problems presented in the previous section.

Formally, a tree \(\T = (\N(\T), \mathcal E(\T))\) is a connected acyclic graph defined by a set of nodes \(\N(\T)\) and a set of edges  \(\mathcal E (\T)\), where the edges in \(\mathcal E(\T)\) describe the nodal interconnections~\cite{Bondy1976}.
Trees have been ubiquitously employed in the past in order to efficiently represent operating environments in
robotics applications.
A small sample of works that consider the use of multi-resolution trees in robotics include~\cite{Hauer2015,Hornung2013,Einhorn2011,Kambhampati1986,Hauer2016,Kraetzschmar2004}.
In these works, environment abstractions in the form of multi-resolution hierarchical trees (specifically, quadtrees and octrees) are employed in order to reduce the computational complexity of graph-search algorithms for autonomous planning, or to ease the on-board memory requirements to store environment representations.
While tree structures are a general hierarchical data structure, we will focus our attention in this chapter on environment abstractions in the form of multi-resolution quadtree representations.
It should be noted, however, that the methods discussed and presented in this chapter are applicable to general tree structures, beyond those of quadtrees.

\begin{figure}[t]
	\begin{adjustbox}{max size={0.98\textwidth}}
	\begin{tikzpicture}[scale=0.7,level distance=1.2cm,
			level 1/.style={sibling distance=4.5cm},
			level 2/.style={sibling distance=0.9cm}]

			\node[fill=black, shape = circle, draw, line width = 1pt, minimum size = 2.5mm, inner sep = 0mm] (root) at (0, 0){}
			child {node[fill=black, shape = circle, draw, line width = 1pt, minimum size = 2.5mm, inner sep = 0mm] (c1) {}
				child {node[shape = circle, draw, line width = 1pt, minimum size = 2.5mm, inner sep = 0mm] (c11) {}}
				child {node[shape = circle, draw, line width = 1pt, minimum size = 2.5mm, inner sep = 0mm] (c12) {}}
				child {node[shape = circle, draw, line width = 1pt, minimum size = 2.5mm, inner sep = 0mm] (c13) {}}
				child {node[shape = circle, draw, line width = 1pt, minimum size = 2.5mm, inner sep = 0mm] (c14) {}}
			}
			child {node[fill=black, shape = circle, draw, line width = 1pt, minimum size = 2.5mm, inner sep = 0mm] (c2) {}
				child {node[shape = circle, draw, line width = 1pt, minimum size = 2.5mm, inner sep = 0mm] (c21) {}}
				child {node[shape = circle, draw, line width = 1pt, minimum size = 2.5mm, inner sep = 0mm] (c22) {}}
				child {node[shape = circle, draw, line width = 1pt, minimum size = 2.5mm, inner sep = 0mm] (c23) {}}
				child {node[shape = circle, draw, line width = 1pt, minimum size = 2.5mm, inner sep = 0mm] (c24) {}}
			}
			child {node[fill=black, shape = circle, draw, line width = 1pt, minimum size = 2.5mm, inner sep = 0mm] (c3) {}
				child {node[shape = circle, draw, line width = 1pt, minimum size = 2.5mm, inner sep = 0mm] (c31) {}}
				child {node[shape = circle, draw, line width = 1pt, minimum size = 2.5mm, inner sep = 0mm] (c32) {}}
				child {node[shape = circle, draw, line width = 1pt, minimum size = 2.5mm, inner sep = 0mm] (c33) {}}
				child {node[shape = circle, draw, line width = 1pt, minimum size = 2.5mm, inner sep = 0mm] (c34) {}}
			}
			child {node[fill=black, shape = circle, draw, line width = 1pt, minimum size = 2.5mm, inner sep = 0mm] (c4) {}
				child {node[shape = circle, draw, line width = 1pt, minimum size = 2.5mm, inner sep = 0mm] (c41) {}}
				child {node[shape = circle, draw, line width = 1pt, minimum size = 2.5mm, inner sep = 0mm] (c42) {}}
				child {node[shape = circle, draw, line width = 1pt, minimum size = 2.5mm, inner sep = 0mm] (c43) {}}
				child {node[shape = circle, draw, line width = 1pt, minimum size = 2.5mm, inner sep = 0mm] (c44) {}}
			};

		\node (x1) at ($(c11.south) + (0,-0.4)$) {\(x_1\)};
		\node (x2) at ($(c12.south) + (0,-0.4)$) {\(x_2\)};
		\node (x3) at ($(c13.south) + (0,-0.4)$) {\(x_3\)};
		\node (x4) at ($(c14.south) + (0,-0.4)$) {\(x_4\)};

		\node (x5) at ($(c21.south) + (0,-0.4)$) {\(x_5\)};
		\node (x6) at ($(c22.south) + (0,-0.4)$) {\(x_6\)};
		\node (x7) at ($(c23.south) + (0,-0.4)$) {\(x_7\)};
		\node (x8) at ($(c24.south) + (0,-0.4)$) {\(x_8\)};

		\node (x9) at ($(c31.south) + (0,-0.4)$) {\(x_9\)};
		\node (x10) at ($(c32.south) + (0,-0.4)$) {\(x_{10}\)};
		\node (x11) at ($(c33.south) + (0,-0.4)$) {\(x_{11}\)};
		\node (x12) at ($(c34.south) + (0,-0.4)$) {\(x_{12}\)};

		\node (x13) at ($(c41.south) + (0,-0.4)$) {\(x_{13}\)};
		\node (x14) at ($(c42.south) + (0,-0.4)$) {\(x_{14}\)};
		\node (x15) at ($(c43.south) + (0,-0.4)$) {\(x_{15}\)};
		\node (x15) at ($(c44.south) + (0,-0.4)$) {\(x_{16}\)};

		\node (treeLabel) at ($(root.north) + (0.4,0.4)$) {\large \(\T_\W\)};
		\end{tikzpicture}
		\hfil
		\begin{tikzpicture}[scale=0.85,every node/.style={minimum size=1cm}]

			\draw[step=1cm, black] (0,0) grid (4,4);

			\node (x1) at (0.5,0.5) {\(x_1\)};
			\node (x2) at (1.5,0.5) {\(x_2\)};
			\node (x3) at (0.5,1.5) {\(x_3\)};
			\node (x4) at (1.5,1.5) {\(x_4\)};

			\node (x5) at (2.5,0.5) {\(x_5\)};
			\node (x6) at (3.5,0.5) {\(x_6\)};
			\node (x7) at (2.5,1.5) {\(x_7\)};
			\node (x8) at (3.5,1.5) {\(x_8\)};

			\node (x9) at (0.5,2.5) {\(x_9\)};
			\node (x10) at (1.5,2.5) {\(x_{10}\)};
			\node (x11) at (0.5,3.5) {\(x_{11}\)};
			\node (x12) at (1.5,3.5) {\(x_{12}\)};

			\node (x13) at (2.5,2.5) {\(x_{13}\)};
			\node (x14) at (3.5,2.5) {\(x_{14}\)};
			\node (x15) at (2.5,3.5) {\(x_{15}\)};
			\node (x16) at (3.5,3.5) {\(x_{16}\)};
		\end{tikzpicture}
	\end{adjustbox}

	\vspace{10pt}

	\begin{adjustbox}{max size={0.98\textwidth}}
		\begin{tikzpicture}[scale=0.7, level distance=1.2cm,
			level 1/.style={sibling distance=4.5cm},
			level 2/.style={sibling distance=0.9cm}]

			\node[fill=black, shape = circle, draw, line width = 1pt, minimum size = 2.5mm, inner sep = 0mm] (root) at (0, 0){}
			child {node[fill=black, shape = circle, draw, line width = 1pt, minimum size = 2.5mm, inner sep = 0mm] (c1) {}
				child {node[shape = circle, draw, line width = 1pt, minimum size = 2.5mm, inner sep = 0mm] (c11) {}}
				child {node[shape = circle, draw, line width = 1pt, minimum size = 2.5mm, inner sep = 0mm] (c12) {}}
				child {node[shape = circle, draw, line width = 1pt, minimum size = 2.5mm, inner sep = 0mm] (c13) {}}
				child {node[shape = circle, draw, line width = 1pt, minimum size = 2.5mm, inner sep = 0mm] (c14) {}}
			}
			child {node[shape = circle, draw, line width = 1pt, minimum size = 2.5mm, inner sep = 0mm] (c2) {}
				child[black!15!] {node[shape = circle, draw, line width = 1pt, minimum size = 2.5mm, inner sep = 0mm] (c21) {}}
				child[black!15!] {node[shape = circle, draw, line width = 1pt, minimum size = 2.5mm, inner sep = 0mm] (c22) {}}
				child[black!15!] {node[shape = circle, draw, line width = 1pt, minimum size = 2.5mm, inner sep = 0mm] (c23) {}}
				child[black!15!] {node[shape = circle, draw, line width = 1pt, minimum size = 2.5mm, inner sep = 0mm] (c24) {}}
			}
			child {node[shape = circle, draw, line width = 1pt, minimum size = 2.5mm, inner sep = 0mm] (c3) {}
				child[black!15!] {node[shape = circle, draw, line width = 1pt, minimum size = 2.5mm, inner sep = 0mm] (c31) {}}
				child[black!15!] {node[shape = circle, draw, line width = 1pt, minimum size = 2.5mm, inner sep = 0mm] (c32) {}}
				child[black!15!] {node[shape = circle, draw, line width = 1pt, minimum size = 2.5mm, inner sep = 0mm] (c33) {}}
				child[black!15!] {node[shape = circle, draw, line width = 1pt, minimum size = 2.5mm, inner sep = 0mm] (c34) {}}
			}
			child {node[fill=black, shape = circle, draw, line width = 1pt, minimum size = 2.5mm, inner sep = 0mm] (c4) {}
				child {node[shape = circle, draw, line width = 1pt, minimum size = 2.5mm, inner sep = 0mm] (c41) {}}
				child {node[shape = circle, draw, line width = 1pt, minimum size = 2.5mm, inner sep = 0mm] (c42) {}}
				child {node[shape = circle, draw, line width = 1pt, minimum size = 2.5mm, inner sep = 0mm] (c43) {}}
				child {node[shape = circle, draw, line width = 1pt, minimum size = 2.5mm, inner sep = 0mm] (c44) {}}
			};

			\node (t1) at ($(c11.south) + (0,-0.4)$) {\(t_1\)};
			\node (t2) at ($(c12.south) + (0,-0.4)$) {\(t_2\)};
			\node (t3) at ($(c13.south) + (0,-0.4)$) {\(t_3\)};
			\node (t4) at ($(c14.south) + (0,-0.4)$) {\(t_4\)};

			\node[black!15!] (x5) at ($(c21.south) + (0,-0.4)$) {\(x_5\)};
			\node[black!15!] (x6) at ($(c22.south) + (0,-0.4)$) {\(x_6\)};
			\node[black!15!] (x7) at ($(c23.south) + (0,-0.4)$) {\(x_7\)};
			\node[black!15!] (x8) at ($(c24.south) + (0,-0.4)$) {\(x_8\)};

			\node[black!15!] (x9) at ($(c31.south) + (0,-0.4)$) {\(x_9\)};
			\node[black!15!] (x10) at ($(c32.south) + (0,-0.4)$) {\(x_{10}\)};
			\node[black!15!] (x11) at ($(c33.south) + (0,-0.4)$) {\(x_{11}\)};
			\node[black!15!] (x12) at ($(c34.south) + (0,-0.4)$) {\(x_{12}\)};

			\node (t7) at ($(c41.south) + (0,-0.4)$) {\(t_{7}\)};
			\node (t8) at ($(c42.south) + (0,-0.4)$) {\(t_{8}\)};
			\node (t9) at ($(c43.south) + (0,-0.4)$) {\(t_{9}\)};
			\node (t10) at ($(c44.south) + (0,-0.4)$) {\(t_{10}\)};

			\node (t5) at ($(c2.west) + (-0.4,0)$) {\(t_{5}\)};
			\node (t6) at ($(c3.east) + (0.4,0)$) {\(t_{6}\)};

			\node (treeLabel) at ($(root.north) + (0.4,0.4)$) {\large \(\T\)};
		\end{tikzpicture}
		\hfil
		\begin{tikzpicture}[scale=0.85,every node/.style={minimum size=1cm}]

		\draw[step=1cm, black] (0,0) grid (2,2); 
		\draw[step=1cm, black] (2,2) grid (4,4); 
		\draw[step=2cm, black] (2,0) grid (4,2); 
		\draw[step=2cm, black] (0,2) grid (2,4); 

		\node (t1) at (0.5,0.5) {\(t_1\)};
		\node (t2) at (1.5,0.5) {\(t_2\)};
		\node (t3) at (0.5,1.5) {\(t_3\)};
		\node (t4) at (1.5,1.5) {\(t_4\)};

		\node (t7) at (2.5,2.5) {\(t_{7}\)};
		\node (t8) at (3.5,2.5) {\(t_{8}\)};
		\node (t8) at (2.5,3.5) {\(t_{9}\)};
		\node (t10) at (3.5,3.5) {\(t_{10}\)};

		\node (t5) at (3,1) {\(t_{5}\)};
		\node (t6) at (1,3) {\(t_{6}\)};
	\end{tikzpicture}
	\end{adjustbox}

	\caption{Visual depiction of the tree \(\T_\W \in \T^\Q\) along with a tree \(\T \in \T^\Q\) and corresponding multi-resolution (grid) representations of \(\W\). Elements of the set \(\Nint(\cdot)\) are shown in black, whereas elements of the set \(\Nlf(\cdot)\) are white and numbered. 
	Notice that the tree \(\T\) aggregates nodes \(x_5,~x_6,~x_7,~\text{and}~ x_8\) to \(t_5\) and the nodes \(x_5,~x_6,~x_7,~\text{and}~ x_8\) to \(t_6\). 
	The nodes \(x_1,~x_2,~x_3,~\text{and}~x_4\) as well as \(x_{13},~x_{14},~x_{15},~\text{and}~x_{16}\) are not aggregated in this example. Aggregated nodes of \(\T_\W\) are shown in gray.}
	\label{fig:encoder_tree_equivalence}
\end{figure}
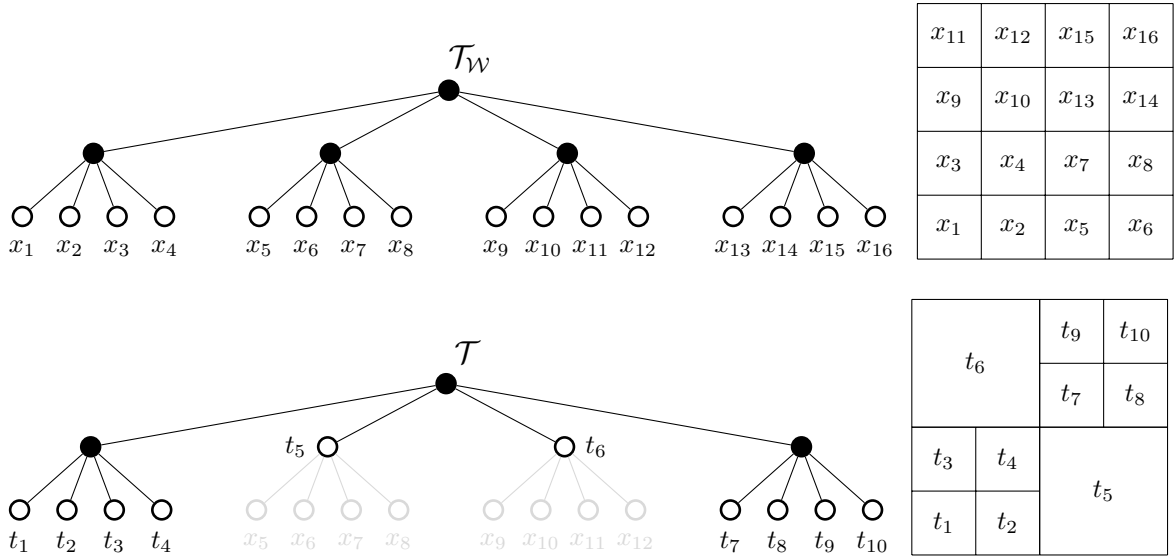

To formalize the connection between an environment and hierarchical tree structures, we will assume that there exists an integer \(\ell > 0\) such that the world \(\W \subset \Re^2\) (generalizable to \(\Re^n\)), is contained within a square (hypercube) of side length \(2^{\ell}\).
The environment \(\W\) is assumed to be a two-dimensional grid-world where each element of \(\W\) is a square (hypercube) of unit side length.
We will take \(\T^{\Q}\) to denote the set of all valid, multi-resolution, quadtree representations of \(\W\), and we let \(\T_\W \in \T^\Q\) be the finest resolution tree of the world \(\W\).
An illustrative example of \(\T_\W\) for a \(4 \times 4\) world \(\W\) is shown in Fig.~\ref{fig:encoder_tree_equivalence}.
Specific terminology is used to refer to some of the inter-node connections in \(\T_\W\), which we describe in the following definition.

\vspace{5pt}

\begin{definition}[\hspace{-0.4pt}\cite{Larsson2020}]
	Let $t \in \mathcal N(\mathcal T_{\mathcal W})$ be any node at depth $k \in \left\{0,\ldots,\ell \right\}$.
	Then $t' \in \mathcal N(\mathcal T_{\mathcal W})$ is a \emph{child} of $t$ if the following hold:
	\begin{enumerate}
		\item
		Node $t'$ is at depth $k+1$ in $\mathcal T_{\mathcal W}$.

		\item
		Nodes $t$ and $t'$ are incident to a common edge, i.e., $\left(t,t'\right) \in \mathcal E \left( \mathcal T_{\mathcal W} \right)$.
	\end{enumerate}
	Conversely, we say that $t$ is the \emph{parent} of $t'$ if $t'$ is a child of $t$.
	Furthermore, we let
	\begin{equation*}
	\mathcal N_k(\mathcal T) = \left\{t \in \mathcal N(\mathcal T) : t \text{~is at depth~} k~\text{in}~\mathcal T_{\mathcal W} \right\},
	\end{equation*}
	to be the set of all nodes of the tree $\mathcal T \in \mathcal T^{\mathcal Q}$ at depth $k$. \hfill $\triangle$
\end{definition}

Accordingly, for any \(t \in \N(\T_\W)\), we let \(\chd(t)\) denote the set of children of the node \(t\).
Furthermore, given any tree \(\T \in \T^\Q\), we say that the node \(t\) is a \emph{leaf node} of \(\mathcal T\) if \(\mathcal C(t) \cap \mathcal N(\mathcal T) = \varnothing\) and denote the set of all leaf nodes for any \(\T \in \T^\Q\) by \(\Nlf(\T) = \{t \in \N(\T_\W): \mathcal C(t) \cap \mathcal N(\mathcal T_{q}) = \emptyset\}\).\footnote{Note that the set \(\N(\T_\W)\) contains all the available nodes.}
The nodes in the set \(\Nint(\T) = \N(\T) \setminus \Nlf(\T)\) are called \emph{interior nodes}.
An example showing the relation between the sets \(\Nlf(\T)\) and \(\Nint(\T)\) is presented in Fig.~\ref{fig:encoder_tree_equivalence}.

The connection between hierarchical data structures, such as multi-resolution quadtrees, and signal encoders \(p(t|x)\) described in Section~\ref{subsec:ITSignalCompression} was recently discussed in~\cite{Larsson2020}.
As part of their study, the authors of~\cite{Larsson2020} discuss how any tree \(\T_q \in \T^\Q\) can be viewed as a deterministic encoder\footnote{A deterministic encoder is an encoder \(p(t|x)\) such that \(p(t|x) \in \{0,1\}\) for all \(t\) and \(x\). 
Deterministic encoders are also called hard encoders.} \(p_q(t|x)\) that maps leaf nodes \(x \in \Nlf(\T_\W)\) to leaf nodes \(t \in \Nlf(\T_q)\) of the tree \(\T_q\).
Specifically, given any tree \(\T_q \in \T^\Q\) a corresponding deterministic encoder \(p_q(t|x)\) can be constructed such that \(p_q(t|x) = 1\) if and only if \(x \in \Nlf(\T_\W)\) is aggregated to the leaf node \(t \in \Nlf(\T_q)\), and  \(p(t|x) = 0\) otherwise.
Consequently, by changing the tree \(\T_q\) we change the corresponding encoder \(p_q(t|x)\), which, in turn, changes the multi-resolution environment representation of \(\W\) defined by the leafs of \(\T_q\), as shown in Fig.~\ref{fig:encoder_tree_equivalence}.
By facilitating a connection between signal encoders and hierarchical trees allows us to utilize the compression frameworks discussed in Section~\ref{subsec:ITSignalCompression} in order to develop problem formulations for the emergence of task-driven, multi-resolution, environment abstractions that can be tailored to agent resource constraints.
We now turn to describe how such a problem formulation can be developed, and the technical considerations that one must observe when searching for a solution to the resulting problem.

\section{Information-Theoretic Multi-Resolution Tree Abstractions}

From the discussion in the previous section, it follows that if we are provided a grid-world \(\W\), then the finest resolution tree corresponding to \(\W\) is given by the tree \(\T_\W\) whose leafs \(x \in \Nlf(\T_\W)\) are the original grid-cells of the environment.
Our objective in this section is to describe how one can utilize the information-theoretic ideas from Section~\ref{subsec:ITSignalCompression} in order to formulate a problem whose solution is a multi-resolution, hierarchical tree representation of the world \(\W\).

To this end, we assume that a grid-world \(\W\) is provided along with a joint distribution \(p(x,y)\).
The joint distribution \(p(x,y)\) captures the correlation between the original grid cells \(X\) and the relevant random variable \(Y\).
For example, if the world \(\W\) is an occupancy grid and the objective is to form abstractions of \(\W\) that are maximally predictive regarding the occupancy values, then \(p(x,y)\) may be specified as \(p(x,y) = p(y|x) p(x)\), where the occupancy information is described by the conditional distribution \(p(y|x)\), and \(p(x)\) is a distribution over world cells (e.g., distribution over possible locations of an autonomous agent).
Since each tree \(\T_q \in \T^{\Q}\) can be represented by a \textit{deterministic} encoder \(p_q(t|x)\), it follows that by changing the tree, we alter the joint distribution \(p_q(t,x,y)\) according to \(p_q(t,x,y) = p_q(t|x) p(x,y)\).
Consequently, a random variable \(T_q: \Omega \to \Re\) can be constructed for each tree \(\T_q \in \T^{\Q}\), whose outcomes are the leafs of \(\T_q\) and has distribution \(p_q(t) = \sum_x p_q(t|x) p(x)\).
The random variable \(T_q\) for each \(\T_q \in \T^\Q\) can be viewed as a compressed representation of the random variable \(X\) created by aggregating finest-resolution grid-cells of the world \(\W\).
Thus, for each tree \(\T_q \in \T^\Q \) we define the mappings \(I_Y : \T^\Q \to [0, \infty)\) and \(I_X : \T^\Q \to [0, \infty)\) according to the rules
\begin{equation} \label{eq:MI_YTree}
	I_Y(\T_q) = I(T_q;Y) = \sum_{t,y} p_q(t,y) \log \frac{p_q(t,y)}{p_q(t) p(y)},
\end{equation}
and
\begin{equation} \label{eq:MI_XTree}
	I_X(\T_q) = I(T_q;X) = \sum_{t,x} p_q(t,x) \log \frac{p_q(t,x)}{p_q(t) p(x)},
\end{equation}
respectively.
We note that in equations \eqref{eq:MI_YTree} and \eqref{eq:MI_XTree}  the distributions \(p(x)\) and \(p(y)\) are not a function of the tree \(\T_q\in\T^\Q\), as they can be obtained by marginalizing the input distribution \(p(x,y)\).
By changing the tree \(\T_q \in \T^\Q\), we change the encoder \(p_q(t|x)\), thereby changing the amount of relevant information captured by the compressed representation \(T_q\) as well as the multi-resolution depiction of \(\W\).

Given the previous connection between encoders and trees, we observe that formalizing the problem of obtaining multi-resolution tree abstractions of \(\W\) that are maximally informative regarding the task-relevant variable \(Y\), can be done by employing an information-theoretic framework for signal compression with additional constraints on the structure of the encoder \(p(t|x)\).
The additional constraints placed on the encoder structure are necessary to ensure that the resulting solution to the compression problem corresponds to some tree \(\T \in \T^\Q\).
The problem we consider in the remainder of this chapter is now formally stated as follows.
\begin{problem} \label{prob:IBtreeOptimProblem1}
	\textit{Given a joint distribution \(p(x,y)\), a scalar \(\hat D \geq 0\), and the world \(\W\), we consider the problem}
	\begin{equation} \label{eq:IBtreeOptim1}
		\min_{\T_q \in \T^\Q} I_X(\T_q),
	\end{equation}
	\textit{subject to the constraint}
	\begin{equation} \label{eq:IBtreeCons1}
		I_Y(\T_q) \geq \hat{D}.
	\end{equation}
	\hfill $\triangle$
\end{problem}

Problem~\ref{prob:IBtreeOptimProblem1} is equivalent to the problem~\eqref{eq:IBproblemMinOpt} subject to~\eqref{eq:IBproblemMinOptCons} with the added constraint \(\T_q \in \T^\Q\) that the encoder corresponds to a valid tree representation of \(\W\).
Imposing the constraint \(\T_q \in \T^\Q\) introduces a host of additional challenges, as existing methods for obtaining a solution to the problem~\eqref{eq:IBproblemMinOpt}-\eqref{eq:IBproblemMinOptCons} do not consider structural constraints on the encoder \(p(t|x)\).
In fact, the solution to the IB problem is, in general, a stochastic mapping \(p(t|x)\)~\cite{Tishby1999}.
Other approaches, such as the agglomerative information bottleneck (AIB)~\cite{Slonim2000,Slonim2002}, design deterministic encoders by utilizing the IB principle and iteratively merging elements of the compressed representation \(T\) until only the trivial abstraction, where \(T\) only has a single outcome, remains.
Although the end result of AIB is a family of deterministic encoders, the method does not consider the constraints requiring the encoder \(p(t|x)\) to be a member of some space, for example correspond to some \(\T \in \T^\Q\).
Our problem is fundamentally different: we employ the IB principle to obtain a tree \(\T_q \in \T^\Q\), which is a solution to~\eqref{eq:IBtreeOptim1}-\eqref{eq:IBtreeCons1} as a function of \(\hat D \geq 0\), by enforcing the constraint \(\T_q \in \T^\Q\).
Notice that the Problem~\ref{prob:IBtreeOptimProblem1}, which aims to maximize compression subject to a lower-bound on the retained relevant information, is equivalent to the following problem~\cite{GiladBachrach2003}.

\begin{problem} \label{prob:IBtreeOptimProblem2}
	\textit{Given the joint distribution \(p(x,y)\), a scalar \(D \geq 0\), and the world \(\W\), we consider the problem}
	\begin{equation} \label{eq:IBtreeOptim2}
		\max_{\T_q \in \T^\Q} I_Y(\T_q),
	\end{equation}
	\textit{subject to the constraint}
	\begin{equation} \label{eq:IBtreeCons2}
		I_X(\T_q) \leq D.
	\end{equation}
	\hfill $\triangle$
\end{problem}

As opposed to maximizing compression subject to a lower bound on the retained relevant information, Problem~\ref{prob:IBtreeOptimProblem2} aims to maximize the relevant information subject to a bound on the achieved compression.
Specifically, the result of solving problem~\eqref{eq:IBtreeOptim1} subject to~\eqref{eq:IBtreeCons1} as a function of \(\hat D \geq 0\) is a multi-resolution tree \(\T_q \in \T^\Q\) that maximizes compression of \(\W\) while ensuring the resulting representation contains at least \(\hat D\) units of relevant information.
Alternatively, Problem~\ref{prob:IBtreeOptimProblem2} allows us to consider the emergence of tree abstractions of \(\W\) that are maximally informative regarding the relevant variable \(Y\) while constraining the allowable compression to at least \(D \geq 0\) units.
As the value of \(D \geq 0\) in~\eqref{eq:IBtreeCons2} increases, one obtains abstractions that are of higher leaf node cardinality, provided that they relay more task-relevant information.
The two formulations are equivalent, and provide different interpretations of the same information-theoretic compression principle~\cite{GiladBachrach2003}.

The most closely related work to our formulation in Problems~\ref{prob:IBtreeOptimProblem1} and~\ref{prob:IBtreeOptimProblem2} is the previous work by the authors of~\cite{Larsson2020,larsson2021information}.
In~\cite{Larsson2020} the authors considered the problem
\begin{equation} \label{eq:TROpaperProblemStatement}
	\max_{\T_q \in \T^\Q} I_Y(\T_q) - \frac{1}{\beta} I_X(\T_q),
\end{equation}
as a function of \(\beta > 0\), assuming \(\W\) and \(p(x,y)\) are provided.
The optimization problem~\eqref{eq:TROpaperProblemStatement} is equivalent to the problem~\eqref{eq:IBminProbLagrangian} restricted to the space \(\T_q\in \T^\Q\), as can be seen by multiplying~\eqref{eq:IBminProbLagrangian} by the constant \(-\nicefrac{1}{\beta}\).
The authors of \cite{Larsson2020} develop an algorithm to solve~\eqref{eq:TROpaperProblemStatement} as a function of \(\beta > 0\), termed \textit{Q-tree search}, which utilizes a function similar to Q-functions employed in reinforcement learning, in order to decide which nodes in the tree to prune~\cite{Bertsekas2005,SuttonBarto}.
However, a notable drawback to the Q-tree search algorithm is its dependence on \(\beta > 0\), which is, 
in turn, a function of the distribution \(p(x,y)\).
Consequently, selecting a value of \(\beta > 0\) for a given level of desired retained relevant information \(I_Y(\T)\) is non-trivial and is further exacerbated when the environment distribution \(p(x,y)\) is time-varying, as is the case in many autonomous system applications.
In contrast, our formulations in Problems~\ref{prob:IBtreeOptimProblem1} and~\ref{prob:IBtreeOptimProblem2} are independent of \(\beta\), instead allowing for the specification of the constraints directly in terms of constants (\(\hat D\) and \(D\), respectively) that do not depend on \(p(x,y)\).
Moreover, we argue that agent resource constraints, such as on-board memory or capacity of communication channels available to the agent, are more naturally specified in terms of constraints of the form of~\eqref{eq:IBtreeCons2} that have an information-theoretic interpretation relating to channel capacity and data-rate, as compared to specifying the same resource constraints in terms of a scalar \(\beta > 0\).

While our formulation is a well-posed optimization problem, it is not obvious how to obtain the solution.
A brute-force solution approach can be employed that exhaustively generates and enumerates all feasible trees in the set \(\T^\Q\).
However, exhaustive search methods do not scale well to large environments, as the set \(\T^\Q\) quickly grows so as to render the computational cost of exhaustively generating each of the trees in \(\T^\Q\) prohibitively expensive.
Instead, alternative solution approaches must be developed that exploit the structural properties of the problem.
In the next section, we show how Problem~\ref{prob:IBtreeOptimProblem1} (equivalently, Problem~\ref{prob:IBtreeOptimProblem2}) can be posed as an integer linear program.

\section{Information-Theoretic Abstractions via Linear Programming} \label{sec:ITabsLP}

In order to formulate a precise mathematical optimization problem, we look to exploit the structural properties of Problem~\ref{prob:IBtreeOptimProblem1}.
To this end, our overall approach is to write the mutual information terms~\eqref{eq:IBtreeOptim1} and~\eqref{eq:IBtreeCons1} in terms of local changes in the tree, and show how the information content in any tree \(\T_q \in \T^\Q\) can be expressed as the sum of all local changes required to obtain \(\T_q\) when starting from the root node of \(\W\).

To begin, we write the information terms~\eqref{eq:IBtreeOptim1} and~\eqref{eq:IBtreeCons1} of any tree \(\T_q \in \T^\Q\) as
\begin{equation} \label{eq:seqenIxTree}
	I_X(\T_q) = I_X(\T_0) + \sum_{i=0}^{q-1} \left[ I_X(\T_{i+1}) - I_X(\T_i) \right],
\end{equation}
and
\begin{equation} \label{eq:seqenIyTree}
	I_Y(\T_q) = I_Y(\T_0) + \sum_{i=0}^{q-1} \left[ I_Y(\T_{i+1}) - I_Y(\T_i) \right],
\end{equation}
respectively.
Relations~\eqref{eq:seqenIxTree} and~\eqref{eq:seqenIyTree} allow us to write the information content of the tree \(\T_q \in \T^\Q\) in terms of a sequence \(\{\T_0,\ldots,\T_{q-1},\T_q\}\), where \(\T_i \in \T^\Q\) for all \(i \in \{0,\ldots,q\}\).
The validity of equations~\eqref{eq:seqenIxTree} and~\eqref{eq:seqenIyTree} does not depend on the specific sequence \(\{\T_0,\ldots,\T_{q-1},\T_q\}\), instead allowing us to express the value of \(I_X(\T_q)\) and \(I_Y(\T_q)\) as a function of other trees \(\T_0,\ldots,\T_{q-1}\), irrespective of how the intermediate trees are selected.
A key observation is that by selecting the intermediate trees \(\T_0,\ldots,\T_{q-1}\) in a specific way, we arrive at tractable expressions for \(I_X(\T_q)\) and \(I_Y(\T_q)\), as follows.

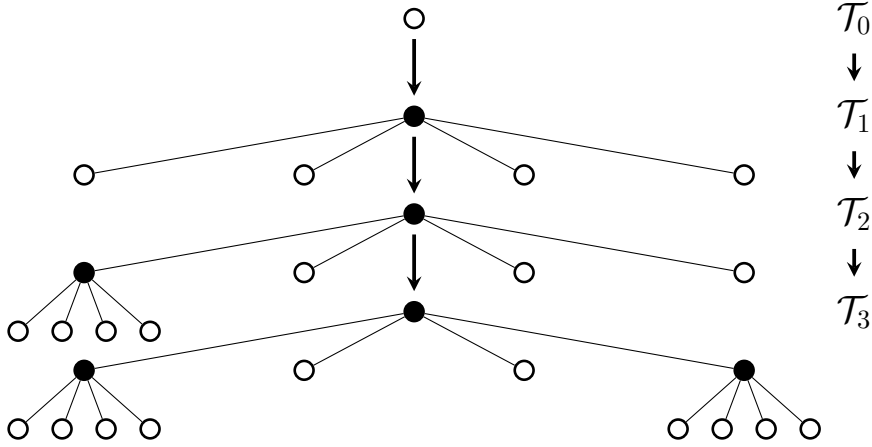
\begin{figure}[t]
	\centering
	\begin{adjustbox}{max size={0.75\textwidth}}
		\begin{tikzpicture}[scale=0.65,level distance=1.2cm,
			level 1/.style={sibling distance=4.5cm},
			level 2/.style={sibling distance=0.9cm}]
			\node[shape = circle, draw, line width = 1pt, minimum size = 2.5mm, inner sep = 0mm] (root1) at (0,0) {};

			\node[fill=black, shape = circle, draw, line width = 1pt, minimum size = 2.5mm, inner sep = 0mm] (root2) at (0,-2) {}
			child {node[shape = circle, draw, line width = 1pt, minimum size = 2.5mm, inner sep = 0mm] {}
			}
			child {node[shape = circle, draw, line width = 1pt, minimum size = 2.5mm, inner sep = 0mm] {}
			}
			child {node[shape = circle, draw, line width = 1pt, minimum size = 2.5mm, inner sep = 0mm] {}
			}
			child {node[shape = circle, draw, line width = 1pt, minimum size = 2.5mm, inner sep = 0mm] {}
			};
			\node[fill=black, shape = circle, draw, line width = 1pt, minimum size = 2.5mm, inner sep = 0mm] (root3) at (0,-4) {}
			child {node[fill=black, shape = circle, draw, line width = 1pt, minimum size = 2.5mm, inner sep = 0mm] {}
				child {node[shape = circle, draw, line width = 1pt, minimum size = 2.5mm, inner sep = 0mm] {}}
				child {node[shape = circle, draw, line width = 1pt, minimum size = 2.5mm, inner sep = 0mm] {}}
				child {node[shape = circle, draw, line width = 1pt, minimum size = 2.5mm, inner sep = 0mm] {}}
				child {node[shape = circle, draw, line width = 1pt, minimum size = 2.5mm, inner sep = 0mm] {}}
			}
			child {node[shape = circle, draw, line width = 1pt, minimum size = 2.5mm, inner sep = 0mm] {}
			}
			child {node[shape = circle, draw, line width = 1pt, minimum size = 2.5mm, inner sep = 0mm] {}
			}
			child {node[shape = circle, draw, line width = 1pt, minimum size = 2.5mm, inner sep = 0mm] {}
			};
			\node[fill=black, shape = circle, draw, line width = 1pt, minimum size = 2.5mm, inner sep = 0mm] (root4) at (0, -6){}
			child {node[fill=black, shape = circle, draw, line width = 1pt, minimum size = 2.5mm, inner sep = 0mm] {}
				child {node[shape = circle, draw, line width = 1pt, minimum size = 2.5mm, inner sep = 0mm] {}}
				child {node[shape = circle, draw, line width = 1pt, minimum size = 2.5mm, inner sep = 0mm] {}}
				child {node[shape = circle, draw, line width = 1pt, minimum size = 2.5mm, inner sep = 0mm] {}}
				child {node[shape = circle, draw, line width = 1pt, minimum size = 2.5mm, inner sep = 0mm] {}}
			}
			child {node[shape = circle, draw, line width = 1pt, minimum size = 2.5mm, inner sep = 0mm] {}
			}
			child {node[shape = circle, draw, line width = 1pt, minimum size = 2.5mm, inner sep = 0mm] {}
			}
			child {node[fill=black, shape = circle, draw, line width = 1pt, minimum size = 2.5mm, inner sep = 0mm] {}
				child {node[shape = circle, draw, line width = 1pt, minimum size = 2.5mm, inner sep = 0mm] {}}
				child {node[shape = circle, draw, line width = 1pt, minimum size = 2.5mm, inner sep = 0mm] {}}
				child {node[shape = circle, draw, line width = 1pt, minimum size = 2.5mm, inner sep = 0mm] {}}
				child {node[shape = circle, draw, line width = 1pt, minimum size = 2.5mm, inner sep = 0mm] {}}
			};
			\draw [->, >=stealth, line width=0.5mm, black] ($(root1.south) + (0,-0.2)$) -- ($(root2.north) + (0,0.2)$);

			\draw [->, >=stealth, line width=0.5mm, black] ($(root2.south) + (0,-0.2)$) -- ($(root3.north) + (0,0.2)$);

			\draw [->, >=stealth, line width=0.5mm, black] ($(root3.south) + (0,-0.2)$) -- ($(root4.north) + (0,0.2)$);
			\node (tree1Label) at ($(root1) + (9,0)$) {\Large \(\mathcal T_0\)}; 
			\node (tree2Label) at ($(root2) + (9,0)$) {\Large \(\mathcal T_1\)}; 
			\node (tree3Label) at ($(root3) + (9,0)$) {\Large \(\mathcal T_2\)}; 
			\node (tree4Label) at ($(root4) + (9,0)$) {\Large \(\mathcal T_3\)}; 

			\draw [->, >=stealth, line width=0.5mm, black] ($(tree1Label.south) + (0,-0.2)$) -- ($(tree2Label.north) + (0,0.2)$); 
			\draw [->, >=stealth, line width=0.5mm, black] ($(tree2Label.south) + (0,-0.2)$) -- ($(tree3Label.north) + (0,0.2)$); 
			\draw [->, >=stealth, line width=0.5mm, black] ($(tree3Label.south) + (0,-0.2)$) -- ($(tree4Label.north) + (0,0.2)$); 
		\end{tikzpicture}
	\end{adjustbox}
	\caption{Sequence of trees \(\left\{ \mathcal T_i \right\}_{i=0}^{3} \subseteq \mathcal T^{\mathcal Q}\) leading from \(\mathcal T_0 = \mathcal R_\W\), the root of \(\T_\W\), to the tree \(\mathcal T_3\).
	Note that \(\mathcal N\left( \mathcal T_{i+1}\right) \setminus \mathcal N\left( \mathcal T_{i}\right) = \mathcal C(t)\) for some \(t \in \Nlf\left( \mathcal T_i \right)\) holds for all \(i \in \{0,1,2\}\). 
	For each tree \(\T_i\), \(i \in \{0,1,2,3\}\), the set of interior nodes, \(\Nint(\T_i)\), is colored black whereas nodes in \(\Nlf(\T_i)\) are white. For the tree \(\mathcal{R}_\W\), the set of interior nodes is empty.}
	\label{fig:sequenceOfTrees}
\end{figure}

To this end, we notice that any tree \(\T_q \in \T^\Q\) can be expressed as a sequence of trees \(\{\T_0,\ldots,\T_{q-1}\}\) by starting at the root tree\footnote{The root tree is the tree \(\R_\W \in \T^\Q\) for which \(\Nint(\R_\W) = \varnothing\). 
That is, \(\R_\W\) consists of only a single leaf node.} \(\T_0 = \R_\W\) and expanding a leaf node of each \(\T_i\), \(i\in\{0,\ldots,q-1\}\), until we reach \(\T_q\).
An illustrative example is shown in Fig.~\ref{fig:sequenceOfTrees}.
By considering sequences of this form, each tree \(\T_i\) in the sequence \(\{\T_i\}_{i=0}^{q-1}\) differs from the tree \(\T_{i+1}\) by only a single leaf node expansion.
In other words, for all \(i \in \{0,\ldots,q-1\}\) it holds that \(\N(\T_{i+1}) \setminus \N(\T_i) = \chd(t)\) for some \(t \in \Nlf(\T_i)\).
We will show that, in this case, the difference in information \(I_X(\T_{i+1}) - I_X(\T_i)\) and \(I_Y(\T_{i+1}) - I_Y(\T_i)\) in~\eqref{eq:seqenIxTree} and~\eqref{eq:seqenIyTree} have a special form.

To elucidate this special form, let us define, for any two trees \(\T_{i+1},\T_i \in \T^\Q\) such that \(\N(\T_{i+1}) \setminus \N(\T_i) = \chd(t)\) for some \(t \in \Nlf(\T_i)\), the function \(\Delta I_X\) 
\begin{equation} \label{eq:deltaIxTrees}
	\Delta I_X(\T_{i+1}, \T_i) = I_X(\T_{i+1}) - I_X(\T_{i}).
\end{equation}
From the definition of the function \(I_X\) in~\eqref{eq:MI_XTree}, as well as the properties of mutual information~\eqref{eq:mutualInfoEntropy}, we see that~\eqref{eq:deltaIxTrees} can be written as
\begin{equation} \label{eq:intermediateDeltaXTree}
	\Delta I_X(\T_{i+1}, \T_i) = H(T_{i+1})- H(T_{i}) - H(T_{i+1} | X) + H(T_i | X).
\end{equation}
Since each tree \(\T_q \in \T^\Q\) defines a deterministic encoder \(p_q(t|x)\) (see Section~\ref{subsec:treesAsEncoders}), it follows that there is no uncertainty in the random variable \(T_q\) when \(X\) is known (i.e., \(X\) specifies the outcome of \(T_q\) with certainty), leading to \(H(T_q | X) = 0\) for all \(\T_q \in \T^\Q\).
As a result, equation~\eqref{eq:intermediateDeltaXTree} becomes
\begin{equation} \label{eq:deltaIxTrees1}
	\Delta I_X(\T_{i+1}, \T_i) = H(T_{i+1})- H(T_{i}).
\end{equation}
Strictly speaking, \eqref{eq:deltaIxTrees1} is valid for any deterministic encoding of \(X\), as we have not yet utilized the fact that \(\N(\T_{i+1}) \setminus \N(\T_i) = \chd(t)\) for some \(t \in \Nlf(\T_i)\) to arrive at this relation.
Now let \(\chd(t) = \{t'_1,\ldots,t'_4\}\) be the set of children of the node \(t \in \Nlf(\T_i)\), and utilize the definition of entropy~\eqref{eq:ShannonEntropy} to write
\begin{equation} \label{eq:entropyExpansionDeltaX}
	\Delta I_X(\T_{i+1}, \T_i) = H(T_{i+1})- H(T_{i}) = - \sum_{i=1}^{4} p(t'_i) \log p(t'_i) + p(t) \log p(t),
\end{equation}
which follows since the only difference between the trees \(\T_i\) and \(\T_{i+1}\) is the aggregation of the nodes in \(\chd(t) \subseteq \Nlf(\T_{i+1})\) to their parent \(t \in \Nlf(\T_i)\), where all other nodes remain unchanged.
Note that 
\begin{equation} \label{eq:treePtDist}
	p(t) = \sum_{i=1}^{4} p(t'_i),
\end{equation}
which follows immediately from the properties of the tree structure and of deterministic encoders.
Using~\eqref{eq:treePtDist}, relation~\eqref{eq:entropyExpansionDeltaX} is equivalently written as
\begin{equation} \label{eq:deltaIxTreeEntropyNode}
	\Delta I_X(\T_{i+1}, \T_i) = p(t) \left[ - \sum_{i=1}^{4} \frac{p(t'_i)}{p(t)} \log \frac{p(t'_i)}{p(t)}\right] = p(t) H(\Pi),
\end{equation}
where \(\Pi \in \Re^4\) is the distribution defined by
\begin{equation} \label{eq:defMergerDistPI}
	\Pi = \left[ \frac{p(t'_1)}{p(t)}, \ldots, \frac{p(t'_4)}{p(t)}\right].
\end{equation}
Equation~\eqref{eq:deltaIxTreeEntropyNode} implies that when the trees \(\T_{i+1}\) and \(\T_i\) differ by only a single leaf node expansion, the change in information \(\Delta I_X(\T_{i+1},\T_i)\) is only a function of the node \(t\in\Nlf(\T_i)\) expanded to create \(\T_{i+1}\), and it does not depend on the other nodes in the tree \(\T_i\).
As a result, while we write \(\Delta I_X(\T_{i+1},\T_i)\) as a function of trees, it can be equivalently written as a function of nodes \(\Delta \hat I_X : \Nint(\T_\W) \to [0,\infty)\) defined by\footnote{To evaluate the function \(\Delta \hat I_X(t)\), we require the node \(t\) to have children. 
The collection of all nodes that have children are those in the set \(\Nint(\T_\W)\).}
\begin{equation} \label{eq:nodeWiseDeltaX}
	\Delta \hat I_X(t) = p(t) H(\Pi).
\end{equation}
Relation~\eqref{eq:nodeWiseDeltaX} quantifies the difference \(\Delta I_X(\T_{i+1},\T_i)\) between \textit{any} two trees \(\T_{i+1},\T_i \in \T^\Q\) that differ only by the expansion of \(t \in \Nlf(\T_i)  \subseteq \Nint(\T_\W)\).

Having obtained a simplified expression for \(\Delta I_X(\T_{i+1},\T_i)\) in~\eqref{eq:nodeWiseDeltaX}, we now turn our attention to \(\Delta I_Y(\T_{i+1},\T_i)\).
In a similar spirit, we define the function \(\Delta I_Y\) between trees \(\T_{i+1}\) and \(\T_i\) such that \(\N(\T_{i+1}) \setminus \N(\T_i) = \chd(t)\) for some \(t \in \Nlf(\T_i)\), according to the rule
\begin{equation} \label{eq:deltaIyTrees1}
	\Delta I_Y(\T_{i+1}, \T_i) = I_Y(\T_{i+1}) - I_Y(\T_i).
\end{equation}
Employing the definition of mutual information~\eqref{eq:mutualInfoDef} and observing that the trees \(\T_{i+1}\) and \(\T_i\) are such that \(\N(\T_{i+1}) \setminus \N(\T_i) = \chd(t)\) for some \(t \in \Nlf(\T_i)\), one obtains
\begin{equation} \label{eq:deltaYinfoTreeIntermediate}
	\Delta I_Y(\T_{i+1}, \T_i) = p(t) \left[ \sum_{i=1}^{4} \frac{p(t'_i)}{p(t)} \sum_y p(y|t'_i) \log p(y|t'_i) - \sum_y p(y|t) \log p(y|t) \right].
\end{equation}
From the tree structure, one can show that for any \(t \in \Nint(\T_\W)\)
\begin{equation} \label{eq:condDistYinfoTree}
	p(y|t) = \sum_{i=1}^{4}[\Pi]_i p(y|t'_i),
\end{equation}
where \(\Pi\) is defined as in~\eqref{eq:defMergerDistPI}.
As a result of~\eqref{eq:deltaYinfoTreeIntermediate} and~\eqref{eq:condDistYinfoTree}, it follows that~\eqref{eq:deltaIyTrees1} can be written as
\begin{equation} \label{eq:deltaIyKLdiv}
	\Delta I_Y(\T_{i+1}, \T_i) = p(t) \sum_{i=1}^{4} [\Pi]_i \mathrm{D}_{\mathrm{KL}} (p(y|t_i), p(y|t)).
\end{equation}
Relation~\eqref{eq:deltaIyKLdiv} is equivalent to
\begin{equation} \label{eq:finalDeltaTreeY}
	\Delta I_Y(\T_{i+1}, \T_i) = p(t) \mathrm{JS}_{\Pi}(p(y|t'_1),\ldots,p(y|t'_{4})),
\end{equation}
where \(\mathrm{JS}_\Pi(p(y|t'_1),\ldots,p(y|t'_{4}))\) is the \textit{Jensen-Shannon (JS) divergence} among the distributions \(p(y|t'_1),\ldots,\) \(p(y|t'_4)\) with weights \(\Pi\)~\cite{Lin1991}.

We note that equation~\eqref{eq:deltaIyKLdiv} is only a function of the nodes that differ between the two trees \(\T_{i+1}\) and \(\T_i\), as was the case for \(\Delta I_X\).
Thus, although \(\Delta I_Y(\T_{i+1},\T_i)\) is written as a function of trees, it can be expressed as a function of nodes according to the mapping \(\Delta \hat I_Y: \Nint(\T_\W) \to [0,\infty)\) defined by
\begin{equation}
	\Delta \hat I_Y(t) = p(t) \mathrm{JS}_{\Pi}(p(y|t'_1),\ldots,p(y|t'_{4})),
\end{equation}
where \(\Pi\) is defined in~\eqref{eq:defMergerDistPI} and \(\chd(t) = \{t'_1,\ldots,t'_4\}\).
Consequently, \(\Delta \hat I_Y(t)\) captures the incremental change in relevant information between \textit{any} two trees \(\T_{i+1}, \T_i \in \T^\Q\) that are such that \(\chd(t) = \N(\T_{i+1}) \setminus \N(\T_i)\) for \(t \in \Nlf(\T_i)\).

Returning to relations~\eqref{eq:seqenIxTree} and~\eqref{eq:seqenIyTree}, we notice that~\eqref{eq:deltaIxTreeEntropyNode} and~\eqref{eq:finalDeltaTreeY} furnish tractable expressions for the terms \(I_X(\T_{i+1}) - I_X(\T_i)\) and \(I_Y(\T_{i+1}) - I_Y(\T_i)\), respectively, when given a sequence of trees \(\{\T_i\}_{i=1}^{q}\) satisfying \(\N(\T_{i+1}) \setminus \N(\T_i) = \chd(t)\) for some \(t \in \Nlf(\T_i)\) and all \(i \in \{0,\ldots,q-1\}\).
Furthermore, from~\eqref{eq:mutualInfoEntropy} and the non-negativity of entropy, we have \(0 \leq I_X(\T_0) \leq H(T_0)\) and \(0 \leq I_Y(\T_0) \leq H(T_0)\).
As a result, selecting the tree \(\T_0\) to be the root tree \(\R_\W\) results in \(I_Y(\T_0) = I_X(\T_0) = 0\), as, in this case, the random variable \(T_0\) has only a single outcome (i.e., \(T_0\) is deterministic), and thus \(H(T_0) = 0\).
Lastly, we observe that for any tree \(\T_q \in \T^\Q\), all sequences \(\{\T_0,\ldots,\T_q\}\) leading from \(\T_0 = \R_\W\) to \(\T_q\) and satisfying \(\N(\T_{i+1}) \setminus \N(\T_i) = \chd(t)\) for some \(t \in \Nlf(\T_i)\) and all \(i \in \{0,\ldots,q-1\}\), must expand all the interior nodes of \(\T_q\).
Accordingly, equations~\eqref{eq:seqenIxTree} and~\eqref{eq:seqenIyTree} can be written, for any \(\T_q \in \T^\Q\), as
\begin{equation} \label{eq:treeXinformationFunction}
	I_X(\mathcal T_q) = \sum_{t \in \Nint(\T_q)} \Delta \hat I_X(t),
\end{equation}
and
\begin{equation} \label{eq:treeYinformationFunction}
	I_Y(\mathcal T_q) = \sum_{t \in \Nint(\T_q)} \Delta \hat I_Y(t),
\end{equation}
respectively.
It should also be noted that relations~\eqref{eq:treeXinformationFunction} and~\eqref{eq:treeYinformationFunction} are valid for all \(\T_q \in \T^\Q\), since every tree in the space \(\T^\Q\) can be obtained by starting at the root tree \(\R_\W \in \T^\Q\) and performing a sequence of expansions according to the process described above.

At this point, we mention two important observations.
First, notice that the quantities \(I_X(\T_q)\) and \(I_Y(\T_q)\) are both fully determined once the interior node set \(\Nint(\T_q)\) is provided.
Secondly, the interior node set also fully characterizes the tree \(\T_q\).
In other words, given the set \(\Nint(\T_q)\), we are able to uniquely determine which tree \(\T_q \in \T^\Q\) corresponds to this interior node set.
As a result, the set \(\Nint(\T_q)\) determines the multi-resolution structure of \(\W\).

\subsection{Integer Linear Program Formulation}

Our goal is  to exploit the previous observations and utilize the expressions~\eqref{eq:treeXinformationFunction} and~\eqref{eq:treeYinformationFunction} to pose Problem~\ref{prob:IBtreeOptimProblem1} as a search for interior nodes.

To this end, we introduce the vector \(\vec{z} \in \Re^{\lvert \Nint(\T_\W) \rvert}\) whose entries \([\vec{z}]_t \in \{0,1\}\) indicate whether or not the node \(t \in \Nint(\T_\W)\) is part of the interior node set \(\Nint(\T_q)\) of the tree \(\T_q\).
The entries of \(\vec{z}\) are such that, for any \(t \in \Nint(\T_\W)\), if \([\vec{z}]_t = 1\) then \(t \in \Nint(\T_q)\), whereas when \([\vec{z}]_t = 0\) implies \(t \notin \Nint(\T_q)\).
We then define the vectors \(\Delta_X \in \Re^{\lvert \Nint(\T_\W) \rvert}\) and \(\Delta_Y \in \Re^{\lvert \Nint(\T_\W) \rvert}\) such that, for all \(t \in \Nint(\T_\W)\), \([\Delta_X]_t = \Delta \hat I_X(t)\) and \([\Delta_Y]_t = \Delta \hat I_Y(t)\).
The relations~\eqref{eq:treeXinformationFunction} and~\eqref{eq:treeYinformationFunction} can then be written in terms of these vectors as \(I_X(\T_q) = \vec{z}^\tp \Delta_X \) and \(I_Y(\T_q) = \vec{z}^{\tp} \Delta_Y\), respectively.
It then follows that Problem~\ref{prob:IBtreeOptimProblem1} can be expressed as the integer linear program (ILP)
\begin{equation} \label{eq:intLinProg2}
    \min_{\z} ~ \z^{\tp} \Delta_X,
\end{equation}
subject to
\begin{align}
    \z^{\tp} \Delta_Y &\geq \hat D, \label{eq:intLinProg2_Cons1}\\
    [\z]_{t'} - [\z]_{t} &\leq 0, ~ t\in \mathcal{B},~ t' \in \chd(t), \label{eq:intLinProg2_Cons2}\\
    [\z]_t &\in \{0,1\}, ~t \in \Nint(\T_\W), \label{eq:intLinProg2_Cons3}
\end{align}
where \(\hat D \geq 0\) and \(\mathcal{B} = \{ t \in \Nint(\T_\W) : \chd(t) \cap \Nint(\T_\W) \neq \varnothing \} \).
The constraint~\eqref{eq:intLinProg2_Cons2} is required as not all vectors \(\vec{z}\) such that \([\vec{z}]_t \in \{0,1\}\) for \(t \in \Nint(\T_\W)\) correspond to a valid tree representation in the space \(\T^\Q\).
Thus, relation~\eqref{eq:intLinProg2_Cons2} ensures that the resulting vector \(\vec{z}\) indeed corresponds to a feasible tree in the space \(\T^\Q\), and does so by enforcing that the children \(t' \in \chd(t)\) of some \(t \in \Nint(\T_\W)\) cannot be expanded unless their parent \(t\) has been 
expanded as well (i.e., if \([\z]_t = 0\) then \([\z]_{t'} = 0\)).
We need only to enforce the constraint~\eqref{eq:intLinProg2_Cons2} for those nodes \(t \in \Nint(\T_\W)\) that have children that are expandable, which are those in the set \(\mathcal B\).

Similarly, it follows that Problem~\ref{prob:IBtreeOptimProblem2} can be formulated as the following ILP
\begin{equation} \label{eq:intLinProg1}
	\max_{\z} ~ \z^{\tp} \Delta_Y,
\end{equation}
subject to the constraints
\begin{align}
	\z^{\tp} \Delta_X &\leq D, \label{eq:intLinProg1_Cons1} \\
	[\z]_{t'} - [\z]_t &\leq 0, ~ t \in \mathcal{B},~ t' \in \chd(t), \label{eq:intLinProg1_Cons2}\\
	[\z]_t &\in \{0,1\}, ~ t \in \Nint(\T_\W), \label{eq:intLinProg1_Cons3}
\end{align}
where \(D \geq 0\) is a given constant.

The two ILPs offer different perspectives of the same information-theoretic principle~\cite{Tishby1999,GiladBachrach2003}.
In the context of resource constrained agents, the formulation~\eqref{eq:intLinProg1}-\eqref{eq:intLinProg1_Cons3} can be viewed as a method for designing multi-resolution tree abstractions for agents that have memory or communication constraints, as the constraint~\eqref{eq:intLinProg1_Cons1} can be interpreted from a data-rate constraint, as discussed in Section~\ref{subsec:ITSignalCompression}.
In these scenarios, the value of \(D\) is chosen so as to reflect the capacity of available communication channels or the amount of available memory resources.
In contrast, the formulation~\eqref{eq:intLinProg2}-\eqref{eq:intLinProg2_Cons3} is for agents who wish to compress the environment representation as much as possible while ensuring at least \(\hat D \geq 0\) units of relevant information are retained in the resulting abstraction.

Our approach allows for a rich set of task-driven, multi-resolution, tree abstractions to be obtained for a given operating environment.
There are, however, many technical considerations pertaining to quantifying the quality of the solutions obtained by solving these programs.
Furthermore, it is not clear how one should compare trees obtained by employing the approach developed in this chapter to existing methods for tree abstraction.
In the next section, we discuss these questions and also introduce an alternative way of obtaining multi-resolution tree abstractions that does not require solving an integer programming problem.

\section{The Information Plane, Pareto Efficiency and LP Relaxation}

This section discusses a number of technical details regarding the solutions to the integer programs formulated in the previous section.
In addition, we also briefly discuss an alternate solution approach to the ILPs \eqref{eq:intLinProg2}-\eqref{eq:intLinProg2_Cons3} and \eqref{eq:intLinProg1}-\eqref{eq:intLinProg1_Cons3} that leverages ideas from convex optimization.
We first consider our formulation in the context of multi-objective optimization theory, investigating whether or not the multi-resolution tree abstractions we obtain by solving the ILPs \eqref{eq:intLinProg2}-\eqref{eq:intLinProg2_Cons3} and \eqref{eq:intLinProg1}-\eqref{eq:intLinProg1_Cons3} are the most informative trees possible for a given level of compression.
To this end, we provide a detailed discussion regarding Pareto efficiency of solutions and introduce a numerical method that can be employed in order to trace the set of Pareto optimal values.
A central component of our discussion of Pareto optimality is the so-called \textit{information plane}, which graphically shows the compression vs. relevant information trade-offs of the solutions to~\eqref{eq:intLinProg2}-\eqref{eq:intLinProg2_Cons3}.
We then close this section by presenting a relaxation approach towards solving the integer programs, which is obtained by removing the integrality constraints~\eqref{eq:intLinProg2_Cons3} and~\eqref{eq:intLinProg1_Cons3}, respectively.
The resulting optimization problem is then a linear program. 
As the solution to the linear program may not be integer, we provide an approach to obtain an integer vector from the linear program solution that is guaranteed to result in a valid multi-resolution tree representation of \(\W\).

\subsection{Pareto Efficiency of Solutions} \label{subsec:ParetoEfficiency}

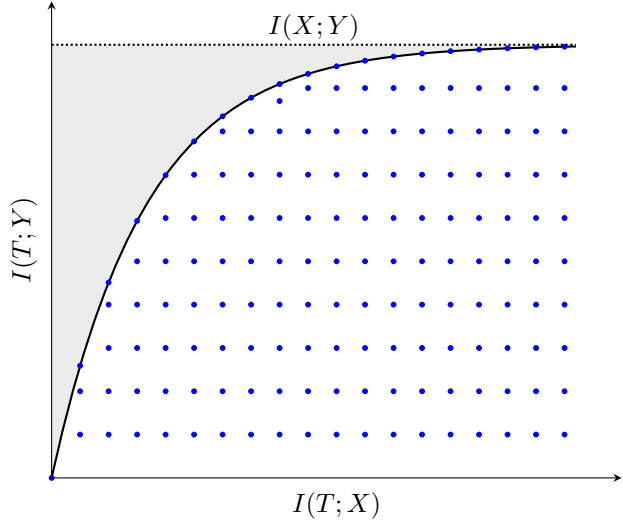
\begin{figure}[tbh]
	\centering
		\begin{tikzpicture}[scale=0.9]
			\begin{axis}[
				axis lines = left,
				xlabel = {$I(T;X)$},
				ylabel = {$I(T;Y)$},
				yticklabels={,,}, xticklabels={,,},
				ytick = \empty, xtick = \empty,
				xmin = 0, xmax = 5,
				ymin = 0, ymax = 1.1,
				clip mode=individual,
				axis on top,
				]

				 \addplot[name path=P, domain=0:4.6, samples=50, color=black, thick]{-exp(-1.2*x) + 1};
	
				 \addplot[name path=I_XY, domain=0:4.6, samples=50, color=black, thick,  densely dotted]{1};
	 
				 \addplot[black,fill opacity=0.08] fill between[of=I_XY and P];

				 \addplot[only marks, blue, mark size = 1pt] coordinates {(0,0)};

				 \addplot[only marks, blue, mark size = 1pt] coordinates {(2,0.87)};

				 \foreach \x in {0.25,0.5,...,4.5}{
				 	\addplot[only marks, blue, mark size = 1pt] coordinates {(\x,{-exp(-1.2*\x)+1})};
					\addplot[only marks, blue, mark size = 1pt] coordinates {(\x,0.1)};
					\addplot[only marks, blue, mark size = 1pt] coordinates {(\x,0.2)};
				 }

				 \foreach \x in {0.5,0.75,...,4.5}{
					\addplot[only marks, blue, mark size = 1pt] coordinates {(\x,0.3)};
					\addplot[only marks, blue, mark size = 1pt] coordinates {(\x,0.4)};
				 }

				 \foreach \x in {0.75,1,...,4.5}{
					\addplot[only marks, blue, mark size = 1pt] coordinates {(\x,0.5)};
				 }
				 
				 \foreach \x in {1,1.25,...,4.5}{
					\addplot[only marks, blue, mark size = 1pt] coordinates {(\x,0.6)};
				 }

				 \foreach \x in {1.25,1.50,...,4.5}{
					\addplot[only marks, blue, mark size = 1pt] coordinates {(\x,0.7)};
				 }

				 \foreach \x in {1.5,1.75,...,4.5}{
					\addplot[only marks, blue, mark size = 1pt] coordinates {(\x,0.8)};
				 }

				 \foreach \x in {2.25,2.5,...,4.5}{
					\addplot[only marks, blue, mark size = 1pt] coordinates {(\x,0.9)};
				 }

				 \node at (axis cs:2.3,1.04) {$I(X;Y)$};
			\end{axis}	
		\end{tikzpicture}
	\caption{Example information plane with the discrete feasible set of trees (blue points). 
	The information plane places relevant information on the ordinate axis and compression on the abscissa (compression decreasing from left to right). The curve traces the trade-off between maximal retention of relevant information vs. compression. In the standard IB problem~\eqref{eq:IBminProbLagrangian}, this curve is characterized by a concave function, but is a discrete set of points in the setting of hierarchical tree abstractions of \(\W\). The grey shaded region above the curve is not achievable by any tree.  The dashed horizontal line is the value of \(I(X;Y)\), which is determined by the input distribution \(p(x,y)\). The solution approaches this value of \(I(X;Y)\) as either \(\beta \to \infty\) in~\eqref{eq:TROpaperProblemStatement} or \(\hat D \to I(X;Y)\) in~\eqref{eq:intLinProg2_Cons1}.}
	\label{fig:informationPlaneAndTrees}
\end{figure}

When discussing the quality of solutions obtained by the IB method, it is common to plot solutions in the so-called information plane.
Specifically, the term \textit{information plane} refers to the graphical display of compression-relevance data on a plot with abscissa \(I(T;X)\) and ordinate \(I(T;Y)\)~\cite{Slonim2002,Tishby1999}.
An example is shown in Fig.~\ref{fig:informationPlaneAndTrees}.

Before we formalize our problem in this section, we discuss a number of important characteristics of the information shown in Fig.~\ref{fig:informationPlaneAndTrees}.
Firstly, it should be noted that the information plane is two-dimensional regardless of the encoder structure.
In other words, the two scalars that quantify the attractiveness of a valid encoder are: (i) its degree of achieved compression \(I(T;X)\), and (ii) the amount of relevant information it retains \(I(T;Y)\).
Secondly, it is important to note that our problem is discrete, as the space of feasible multi-resolution tree representations of \(\W\) is a finite set.
This is in stark contrast to the standard information-bottleneck method, where the decision variables of the optimization problem are the (real-valued) probability mass values of the conditional distribution \(p(t|x)\).
As a result, the curve characterizing the trade-off between the maximal amount of retainable information and compression in the standard information-bottleneck method is a concave function, whereas in our problem it is a discrete set of points.

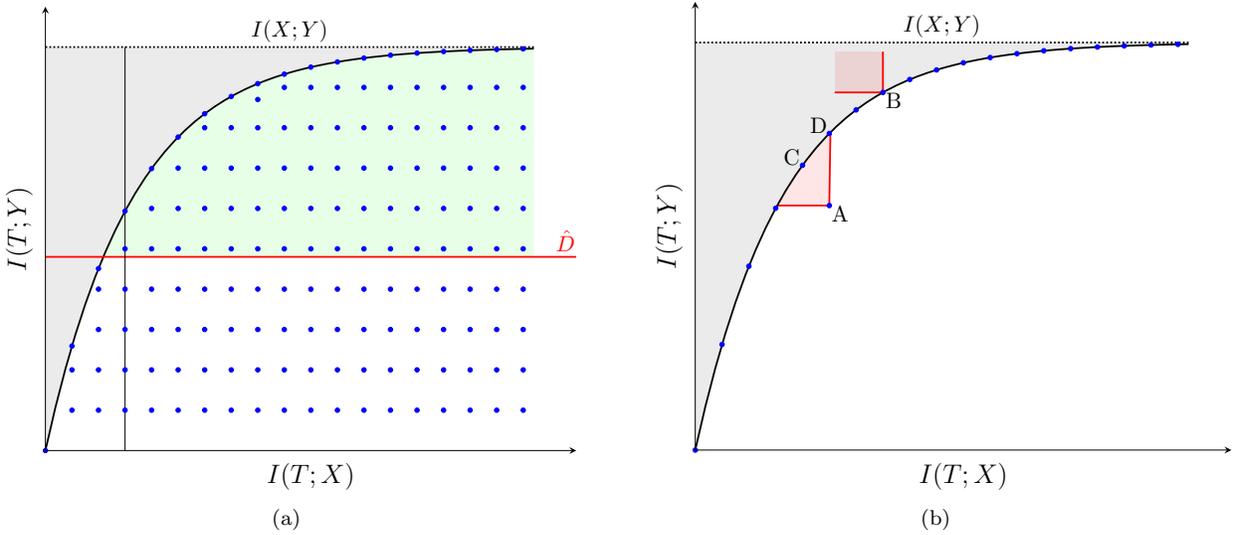
\begin{figure}[tbh]
	\subfloat[]{
		\begin{adjustbox}{max size={0.47\textwidth}}
			\begin{tikzpicture}
				\begin{axis}[
					axis lines = left,
					xlabel = {\large $I(T;X)$},
					ylabel = {\large $I(T;Y)$},
					yticklabels={,,}, xticklabels={,,},
					ytick = \empty, xtick = \empty,
					xmin = 0, xmax = 5,
					ymin = 0, ymax = 1.1,
					clip mode=individual,
					axis on top,
					]

					\addplot[name path=P, domain=0:4.6, samples=50, color=black, thick]{-exp(-1.2*x) + 1};

					\addplot[name path=I_XY, domain=0:4.6, samples=50, color=black, thick, densely dotted]{1};
		
					\addplot[black,fill opacity=0.08] fill between[of=I_XY and P];

					\addplot[name path=C1, domain=0:5, samples=50, color=red, thick]{0.48};
					
					\addplot[name path=V1, domain=0.75, samples=50, color=black, line width=0.01mm]coordinates {(0.75,0)(0.75,1)};

					\addplot [thick, color=green, fill=green, fill opacity=0.1] fill between[of=P and C1, soft clip={domain=0.55:4.6}];
					\addplot[only marks, blue, mark size = 1pt] coordinates {(0,0)};

					\addplot[only marks, blue, mark size = 1pt] coordinates {(2,0.87)};

					\foreach \x in {0.25,0.5,...,4.5}{
						\addplot[only marks, blue, mark size = 1pt] coordinates {(\x,{-exp(-1.2*\x)+1})};
						\addplot[only marks, blue, mark size = 1pt] coordinates {(\x,0.1)};
						\addplot[only marks, blue, mark size = 1pt] coordinates {(\x,0.2)};
					}

					\foreach \x in {0.5,0.75,...,4.5}{
						\addplot[only marks, blue, mark size = 1pt] coordinates {(\x,0.3)};
						\addplot[only marks, blue, mark size = 1pt] coordinates {(\x,0.4)};
					}

					\foreach \x in {0.75,1,...,4.5}{
						\addplot[only marks, blue, mark size = 1pt] coordinates {(\x,0.5)};
					}
					
					\foreach \x in {1,1.25,...,4.5}{
						\addplot[only marks, blue, mark size = 1pt] coordinates {(\x,0.6)};
					}

					\foreach \x in {1.25,1.50,...,4.5}{
						\addplot[only marks, blue, mark size = 1pt] coordinates {(\x,0.7)};
					}

					\foreach \x in {1.5,1.75,...,4.5}{
						\addplot[only marks, blue, mark size = 1pt] coordinates {(\x,0.8)};
					}

					\foreach \x in {2.25,2.5,...,4.5}{
						\addplot[only marks, blue, mark size = 1pt] coordinates {(\x,0.9)};
					}

					\node at (axis cs:2.3,1.04) {$I(X;Y)$};

					\node at (axis cs:4.9,0.52) {\color{red}$\hat D$};

				\end{axis}	
			\end{tikzpicture}
		\end{adjustbox}\label{fig:nonUniquenessOfSoln}}
	\hfill
	\subfloat[]{
		\begin{adjustbox}{max size={0.47\textwidth}}
			\begin{tikzpicture}
				\begin{axis}[
					axis lines = left,
					xlabel = {\large $I(T;X)$},
					ylabel = {\large $I(T;Y)$},
					yticklabels={,,}, xticklabels={,,},
					ytick = \empty, xtick = \empty,
					xmin = 0, xmax = 5,
					ymin = 0, ymax = 1.1,
					clip mode=individual,
					axis on top,
					]
			
					\addplot[name path=L1, domain=0.7635756098951292:1.25, samples=50, color=red, thick]{0.6};
					
					\addplot[name path=L2, domain=1.25, samples=50, color=red, thick]coordinates {(1.25,0.6)(1.26,{-exp(-1.2*1.25)+1})};

					\addplot[only marks, blue, mark size = 1pt] coordinates {(1.25,0.6)};

					\addplot[name path=P, domain=0:4.6, samples=50, color=black, thick]{-exp(-1.2*x) + 1};
		
					\addplot[name path=I_XY, domain=0:4.6, samples=50, color=black, thick, densely dotted]{1};
		
					\addplot[black,fill opacity=0.08] fill between[of=I_XY and P];

					\addplot[name path=L1_2, domain=1.3:1.75, samples=50, color=red, thick]{0.8775435717470181};
					
					\addplot[name path=L2_2, domain=1.75, samples=50, color=red, thick]coordinates {(1.75,{-exp(-1.2*1.75)+1})(1.75,{-exp(-1.2*1.75)+1+0.1})};

					\addplot[name path=dummyL2_2,opacity=0,samples=100,domain=1.3] coordinates{(1.3,{-exp(-1.2*1.75)+1})(1.3,{-exp(-1.2*1.75)+1+0.1})};
					\addplot[only marks, blue, mark size = 1pt] coordinates {(0,0)};

					\foreach \x in {0.25,0.5,...,4.5}{
						\addplot[only marks, blue, mark size = 1pt] coordinates {(\x,{-exp(-1.2*\x)+1})};
					}

					\addplot[thick, color=red, fill=red, fill opacity=0.1] fill between[of=P and L1, soft clip={domain=0.7635756098951292:1.25}];

					\addplot[thick, color=red, fill=red, fill opacity=0.1] fill between[of=L2_2 and dummyL2_2];

					\node at (axis cs:2.3,1.04) {$I(X;Y)$};

					\node at (axis cs:{1.25+0.1},{0.6-0.02}) {A};

					\node at (axis cs:{1.75+0.1},{-exp(-1.2*1.75)+1-0.02}) {B};

					\node at (axis cs:{1-0.1},{-exp(-1.2*1)+1+0.02}) {C};

					\node at (axis cs:{1.25-0.1},{-exp(-1.2*1.25)+1+0.02}) {D};
				\end{axis}	
			\end{tikzpicture}
		\end{adjustbox}\label{fig:paretoSet}}
	\caption{(a) Information plane showing the feasible region \(\{(\z^{\tp}\Delta_X, \z^{\tp}\Delta_Y):\z \in \mathcal D_{\hat D}\}\) (green), and a horizontal line (red) drawn at the given value of \(\hat D\). (b) Set of Pareto optimal values (blue points along black curve) with select solutions (trees) shown.}
	\label{fig:nonUniqunessAndParetoSet}
\end{figure}

In light of the above observations, it is natural to ask whether or not solutions obtained by solving the integer program~\eqref{eq:intLinProg2}-\eqref{eq:intLinProg2_Cons3} are members of the set of trees that characterize the curve in Fig.~\ref{fig:informationPlaneAndTrees}.
Unfortunately, this cannot be guaranteed, in general, as the integer program does not order solutions based on retained relevant information.
To understand why this is the case, consider solving the problem~\eqref{eq:intLinProg2}-\eqref{eq:intLinProg2_Cons3} for a given value of \(\hat D \geq 0\), where the set of feasible solutions is given by
\begin{equation}
    \mathcal D_{\hat D} = \{\z \in \Re^{\lvert \Nint(\T_\W) \rvert}: \eqref{eq:intLinProg2_Cons1}\text{-}\eqref{eq:intLinProg2_Cons3}~\text{are satisfied}\}.
\end{equation}
The feasible set in the information plane for this instance of the problem is shown in Fig.~\ref{fig:nonUniquenessOfSoln}.
Notice that the value of \(\z^{\tp} \Delta_X\) is constant along vertical lines in the information plane. 
As a result, minimizing the cost \(\z^{\tp} \Delta_X\) as in~\eqref{eq:intLinProg2} amounts to finding an element \(\z \in \mathcal D_{\hat D}\) such that the vertical line passing through \(\z\) in the information plane is as close to the origin as possible.
However, as seen in Fig.~\ref{fig:nonUniquenessOfSoln}, there may be multiple feasible trees that minimize \(\z^{\tp} \Delta_X\).
More specifically, we see in the example in Fig.~\ref{fig:nonUniquenessOfSoln} that there are two feasible points along the vertical line intersecting the x-coordinate at the minimum value of \(\z^{\tp}\Delta_X\).
Since the optimization problem~\eqref{eq:intLinProg2}-\eqref{eq:intLinProg2_Cons3} does not order solutions based on their retained relevant information, the two feasible solutions along the vertical line in Fig.~\ref{fig:nonUniquenessOfSoln} are equivalent with respect to the cost function of the integer program~\eqref{eq:intLinProg2}-\eqref{eq:intLinProg2_Cons3}.
Consequently, either one of these two solutions may be returned when executing a suitable search algorithm, and thus the result of the ILP need not be the most informative tree among those that attain the minimum of~\eqref{eq:intLinProg2}.
Despite this, we aim throughout the remainder of this section to develop an approach toward characterizing the set of trees that maximally trade information retention and compression for a given environment \(\W\).
Devising a method for obtaining this set is useful, as it allows for a more meaningful understanding of the overall performance of various abstraction algorithms applied to generate abstractions of \(\W\), and also to elucidate the intrinsic trade-off between information retention and compression of a given environment (i.e., model complexity vs. model quality).
In our endeavor to characterize the set of trees that best trade information retention and compression, we leverage ideas from multi-objective optimization theory, which we turn to discuss next.

As the preceding discussion exemplified, it is insufficient to consider only the degree of achieved compression (\(\z^{\tp}\Delta_X\)) or the amount of retained information (\(\z^{\tp}\Delta_Y\)) as the sole representative of the quality of a candidate solution when searching for solutions that maximize compression and relevant information retention.
Instead, we aim to quantify the goodness of a solution based on \emph{two} measures of quality, namely \emph{both} the compression \emph{and} retained relevant information.
To this end, let the set
\begin{equation}
	\mathcal I_{\hat D} = \{(\z^{\tp}\Delta_X, \z^{\tp}\Delta_Y): \z \in \mathcal D_{\hat D}\},
\end{equation}
be the collection of all feasible compression-relevance pairs for a given value of \(\hat D\).
Our goal is then to select a feasible tree \(\z \in \mathcal D_{\hat D}\) such that the pair \((\z^{\tp}\Delta_X, \z^{\tp}\Delta_Y) \in \mathcal I_{\hat D}\) simultaneously maximizes compression (minimizes \(\z^{\tp}\Delta_X\)) and the retained relevant information (maximize \(\z^{\tp}\Delta_Y\)).
However, searching for an optimal realization of the vector \((\z^{\tp}\Delta_X,\z^{\tp}\Delta_Y) \in \mathcal I_{\hat D}\) over \(\z \in \mathcal D_{\hat D}\) is a multi-objective optimization problem, where feasible solutions \(\z \in \mathcal{D}_{\hat D}\) are not totally ordered with respect to the vector-valued objective \((\z^{\tp} \Delta_X, \z^{\tp}\Delta_Y)\)~\cite{Boyd2004convex}.
For example, consider two candidate solutions \(\z,\hat\z \in \mathcal{D}_{\hat D}\) where \(\z^{\tp} \Delta_X < \hat \z^{\tp} \Delta_X\) and \(\z^{\tp}\Delta_Y < \hat \z^{\tp}\Delta_Y\).
In this scenario, the solution \(\z\) achieves a greater degree of compression but retains less information as compared to \(\hat \z\).
As a result, the solutions \(\z\) and \(\hat \z\) are incomparable, since which of these solutions we prefer depends on our relative preference of compression and information retention.

Being that solutions in multi-objective optimization problems are generally partially ordered, the goal of multi-objective problems must be modified in comparison to their single (scalar) objective optimization counterparts~\cite{Boyd2004convex}.
To this end, methods for multi-objective optimization problems aim to obtain solutions that are \textit{Pareto efficient}~\cite{Boyd2004convex,Das1997}.
In our problem setting, a feasible tree \(\z \in \mathcal D_{\hat D}\) is Pareto efficient if there does not exist a \(\hat \z \in \mathcal D_{\hat D}\) such that \((\hat \z^{\tp}\Delta_X, -\hat\z^{\tp}\Delta_Y) \leq (\z^{\tp}\Delta_X, -\z^{\tp}\Delta_Y)\) holds with strict inequality in at least one component (i.e., either \(\hat\z^{\tp}\Delta_X < \z^{\tp}\Delta_X\) or \(-\hat\z^{\tp}\Delta_Y < -\z^{\tp}\Delta_Y\), or both).
The collection of vectors \((\z^{\tp}\Delta_X, \z^{\tp}\Delta_Y) \in \Re^2\) for all Pareto efficient solutions \(\z \in \mathcal D_{\hat D}\) is called the \textit{set of Pareto optimal values}~\cite{Boyd2004convex}.
An example is shown in Fig.~\ref{fig:paretoSet}.
From Fig.~\ref{fig:paretoSet}, we observe that trees that have information plane points in the red region offer improvement in at least one attribute (compression or information retention) as compared to the solution at the intersection of the red lines. 
For example, the tree \(\textrm{A}\) offers less compression and less retained information than the tree \(\textrm{C}\), and the same compression, but less information retained, as compared with tree \(\textrm{D}\). 
Notice that Pareto efficient trees, such as the tree corresponding to \(\textrm{B}\), do not have any feasible points in the red region other than themselves and possibly other solutions (trees) that are equivalent in the information plane (i.e., trees that are the same in terms of compression and information retention). 
Trees corresponding to points \(\textrm{B}\), \(\textrm{C}\), and \(\textrm{D}\) are all Pareto efficient and are not comparable.

To formalize the notion of Pareto efficiency consider any \(\z \in \mathcal D_{\hat D}\), and define the set
\begin{equation} \label{eq:domSet}
	\mathcal S_{\hat D}(\z) = \{ \hat\z \in \mathcal D_{\hat D} : \hat\z^{\tp}\Delta_Y \geq \z^{\tp} \Delta_Y ~~\text{and}~~ \hat\z^{\tp}\Delta_X \leq \z^{\tp} \Delta_X \}.
\end{equation}
For any tree \(\z \in \mathcal D_{\hat D}\), the set \(\mathcal S_{\hat D}(\z)\) contains those trees \(\hat \z\) that satisfy any one of the following conditions: (i) contain the same amount of relevant information for the given amount of compression as compared to the tree \(\z\), (ii) provide more relevant information for the same level of compression, (iii) achieve a greater degree of compression for the same amount of relevant information as the tree \(\z\), or (iv) provide both more relevant information and greater degree of compression as compared to the tree \(\z\).
Also, notice that the set \(\mathcal S_{\hat D}(\z)\) contains those trees that have the same amount of relevant information and achieve an identical degree of compression (i.e., share a common point in the information plane).
To identify the trees that share a common point in the information plane, we define for any \(\z \in \mathcal D_{\hat D}\) the set
\begin{equation} \label{eq:setOfEquivalentTrees}
	\mathcal U_{\hat D}(\z) = \{ \hat\z \in\mathcal D_{\hat D} :  \hat\z^{\tp} \Delta_X = \z^{\tp} \Delta_X ~~\text{and}~~ \hat\z^{\tp} \Delta_Y = \z^{\tp} \Delta_Y\}.
\end{equation}
Thus, the trees \(\hat \z \in \mathcal S_{\hat D}(\z) \setminus \mathcal U_{\hat D}(\z)\) achieve a greater degree of compression or retain more relevant information (or both), while being no worse in either attribute as compared to \(\z \in \mathcal D_{\hat D}\).
The set of points \(\z \in \mathcal D_{\hat D}\) for which there does not exist any tree \(\hat \z \in \mathcal D_{\hat D}\) that offers improvement in either compression or information retention, while being no worse in either attribute, are those trees in the set 
\begin{equation} \label{eq:paretoFrontier_hatD}
	\mathcal P_{\hat D} = \{\z \in \mathcal D_{\hat D} : \mathcal S_{\hat D}(\z) \setminus \mathcal U_{\hat D}(\z) = \varnothing \}.
\end{equation}
Consequently, the collection of all Pareto efficient solutions is given by 
\begin{equation} \label{eq:wholeParetoFrontier}
	\mathcal P = \{ \z : \z \in \mathcal P_{\hat D}, ~\hat D \geq 0\},
\end{equation}
and the corresponding set of Pareto optimal values is
\begin{equation}
	\mathcal V = \{ (\z^{\tp}\Delta_X, \z^{\tp}\Delta_Y) : \z \in \mathcal P \}.
\end{equation}
Given the mathematical description of the set of Pareto optimal values, we are now interested in the computational aspects of obtaining the set \(\mathcal V\).
Interestingly, obtaining Pareto efficient trees and tracing the set of Pareto optimal values for our problem can be done in a principled manner, as follows.

\subsection{Obtaining Pareto Efficient Trees}

Given value of \(\hat D \geq 0\), consider the optimization problem
\begin{equation} \label{eq:paretoILP1}
    D^* \triangleq \min_{\z} ~ \z^{\tp} \Delta_X,
\end{equation}
subject to
\begin{align}
    \z^{\tp} \Delta_Y &\geq \hat D, \label{eq:paretoILP1_cons1}\\
    [\z]_{t'} - [\z]_{t} &\leq 0, ~ t\in \mathcal{B},~ t' \in \chd(t), \label{eq:paretoILP1_cons2}\\
    [\z]_t &\in \{0,1\}, ~t \in \Nint(\T_\W), \label{eq:paretoILP1_cons3}
\end{align}
which is the same as~\eqref{eq:intLinProg2}-\eqref{eq:intLinProg2_Cons3} and repeated here for convenience.
As discussed previously,  the solution to the above optimization problem is not necessarily a Pareto efficient solution, as there may be a more informative tree for the same level of compression available in the feasible set \(\mathcal D_{\hat D}\).
We are, however, guaranteed that there are no trees containing at least \(\hat D\) units of relevant information that offer more compression than \(D^*\) units. 
Thus, in order to obtain a Pareto efficient tree, we must check to see if there are any other solutions that contain more relevant information with \(D^*\) units of compression.
To this end, we consider the following equality constrained integer program 
\begin{equation} \label{eq:paretoILP2}
	\hat D^* \triangleq \max_{\z} ~ \z^{\tp} \Delta_Y,
\end{equation}
subject to the constraints
\begin{align}
	\z^{\tp} \Delta_X &= D^*, \label{eq:paretoILP2_cons1} \\
	[\z]_{t'} - [\z]_t &\leq 0, ~ t \in \mathcal{B},~ t' \in \chd(t), \label{eq:paretoILP2_cons2}\\
	[\z]_t &\in \{0,1\}, ~ t \in \Nint(\T_\W). \label{eq:paretoILP2_cons3}
\end{align}
It is important to observe that the feasible set of the above linear program is not empty, since the existence of a tree \(\z \in \mathcal D_{\hat D}\) for which \(\z^{\tp} \Delta_X = D^*\) is known from the solution of~\eqref{eq:intLinProg2}-\eqref{eq:intLinProg2_Cons3}.
The resulting pair \((D^*,\hat D^*) \in \mathcal V\) is in the set of Pareto optimal values, and the solution to the problem~\eqref{eq:paretoILP2}-\eqref{eq:paretoILP2_cons3} is a member of the set \(\mathcal P\).

One can then obtain the entire set of Pareto optimal values by performing the aforementioned process for \(\hat D \in [0,I(X;Y)]\).
Having a method for obtaining the set of Pareto optimal values, we are able to quantify the solutions from various abstraction methods (e.g., integer programming, Q-tree search, etc.) with respect to trees that offer the best trade-off between compression and retained (relevant) information, as we will show in the numerical example that follows.
However, before we present some numerical results, we briefly discuss an alternate approach to solving the program~\eqref{eq:intLinProg2}-\eqref{eq:intLinProg2_Cons3} that is obtained by removing the integer constraints~\eqref{eq:intLinProg2_Cons3}.

\subsection{Linear Programming Relaxation} \label{subsec:LPRelax}

The optimization problems~\eqref{eq:intLinProg2}-\eqref{eq:intLinProg2_Cons3} and~\eqref{eq:intLinProg1}-\eqref{eq:intLinProg1_Cons3} fall within the realm of combinatorial optimization due to the integer constraint placed on the decision variable \(\z\)~~\cite{Papadimitriou1998combinatorial,Ahuja1988network}.
As a result, standard gradient and sub-gradient based methods for optimization are not applicable due to the discrete nature of the feasible set of solutions.
Instead, one may observe that by removing (or relaxing) the integrality constraints~\eqref{eq:intLinProg2_Cons3} and~\eqref{eq:intLinProg1_Cons3} we obtain a linear program, which can be solved by employing efficient convex optimization algorithms developed in the past decades.
It is then natural to investigate if relaxing the integrality constraints, thereby circumnavigating the discrete, combinatorial aspects of optimization problem, can result in reasonable solutions (i.e., multi-resolution trees) to our problem.
To this end, the \textit{linear programming (LP) relaxation}~\cite{Papadimitriou1998combinatorial,Ahuja1988network} of the problem~\eqref{eq:intLinProg2}-\eqref{eq:intLinProg2_Cons3} is given by

\begin{equation} \label{eq:LPRelax}
	\min_{\z} ~ \z^{\tp} \Delta_X,
\end{equation}
subject to the constraints
\begin{align}
	\z^{\tp} \Delta_Y &\geq \hat D, \label{eq:lpRelax_cons1}\\
    [\z]_{t'} - [\z]_{t} &\leq 0, ~ t\in \mathcal{B},~ t' \in \chd(t), \label{eq:lpRelax_cons2}\\
    0 \leq [\z]_t  &\leq 1, ~t \in \Nint(\T_\W). \label{eq:lpRelax_cons3}
\end{align}
In the problem~\eqref{eq:LPRelax}-\eqref{eq:lpRelax_cons3}, the vector \(\z \in \Re^{\lvert \Nint(\T_\W) \rvert}\) is not restricted to have integer entries, as is the case in the integer programming formulation~\eqref{eq:intLinProg2}-\eqref{eq:intLinProg2_Cons3}.
Consequently, the problem~\eqref{eq:LPRelax} subject to~\eqref{eq:lpRelax_cons1}-\eqref{eq:lpRelax_cons3} is a linear (convex) program for which traditional approaches for optimization (e.g., the simplex method) may be applied in order to obtain a solution~\cite{Papadimitriou1998combinatorial}.
The LP relaxation for the integer problem~\eqref{eq:intLinProg1}-\eqref{eq:intLinProg1_Cons3} can be formulated analogously.

Notice that since the LP relaxation removes the integer constraint~\eqref{eq:intLinProg2_Cons3}, we can no longer guarantee that the solution to the relaxed problem~\eqref{eq:LPRelax}-\eqref{eq:lpRelax_cons3} corresponds to a valid tree representation of \(\W\).
Therefore, we must post-process the solution vector from the LP relaxation in order to obtain a valid multi-resolution tree depiction of the world \(\W\).
To this end, assume \(\z^* \in \Re^{\lvert \Nint(\T_\W)\rvert}\) is a solution to the LP relaxation~\eqref{eq:LPRelax}-\eqref{eq:lpRelax_cons3}, where the entries \([\z^*]_t\) are not necessarily integer.
Now, we construct a multi-resolution tree representation from the vector \(\z^*\) by considering the function \(f_{\delta}: \Re \to \{0,1\}\) defined according to the rule
\begin{equation}
	f_{\delta}(z) =
	\begin{cases}
		1, & ~\text{if }z \geq \delta, \\
		0, & ~\text{otherwise},
	\end{cases}
\end{equation}
and let the \emph{extension of \(f_\delta\) to \(\Re^n\)} be the function \(\bar f_{\delta}:\Re^n \to \{0,1\}^n\) defined by
\begin{equation} \label{eq:extensionTruncationRule}
	\bar f_{\delta}(\z) = (f_{\delta}([\z]_1),\ldots,f_{\delta}([\z]_n)).
\end{equation}
An integer vector \(\tilde \z^*\) can then be constructed from \(\z^*\) according to \(\tilde \z^* = \bar f_{\delta}(\z^*)\).
Observe that creating the vector \(\tilde \z^*\) from \(\z^*\) according to~\eqref{eq:extensionTruncationRule} not only yields an integer vector, but also guarantees that \(\tilde \z^*\) corresponds to a valid tree representation of~\(\W\).
To see why this is the case, we note that the constraint~\eqref{eq:lpRelax_cons2} implies that the values \([\z^*]_{t'} \in \Re\) and \([\z^*]_t \in \Re\) satisfy \([\z^*]_{t'} \leq [\z^*]_t\) for all \(t \in \mathcal B\), \(t' \in \mathcal \chd(t)\).
Then, since the function \(f_{\delta}\) is monotonic, we have \([\tilde \z^*]_{t'} = f_{\delta}([\z^*]_{t'}) \leq f_{\delta}([\z^*]_{t}) = [\tilde \z^*]_t\), and thus \([\tilde \z^*]_{t'} - [\tilde \z^*]_t \leq 0\) holds for all \(t \in \mathcal B\) and \(t' \in \mathcal \chd(t)\).
As a result, the vector \(\tilde \z^* = \bar f_{\delta}(\z^*)\) is integer and represents a valid tree representation, since the constraints \([\tilde \z^*]_{t'} - [\tilde \z^*]_{t} \leq 0,~t \in \mathcal B,~t' \in \mathcal \chd(t)\) and \([\tilde \z^*]_t \in \{0,1\},~t \in \Nint(\T_\W)\) are both satisfied.
However, creating the integer solution vector \(\tilde \z^*\) according to \(\tilde \z^* = \bar f_{\delta}(\z^*)\) from the solution \(\z^*\) to the LP relaxation~\eqref{eq:LPRelax}-\eqref{eq:lpRelax_cons3} does not guarantee that the resulting multi-resolution representation contains at least \(\hat D\) units of relevant information.
In other words, it is not assured that the integer vector \(\tilde \z^* = \bar f_{\delta}(\z^*)\) satisfies the constraint~\eqref{eq:lpRelax_cons1}.
Despite this drawback, we observe that the LP relaxation method described above performs very well in simulation, as we shall see in the numerical examples we present next.

\section{Numerical Examples and Discussion}

In this section, we present two examples that demonstrate the utility of our framework for generating task-relevant abstractions.
For each example, we provide a sample of abstractions obtained by the integer program and discuss the quality of the solutions in the information plane.
In addition, the set of Pareto optimal values for each case is shown, and the integer program formulation developed in this chapter is compared to other approaches for tree abstraction, such as LP relaxation (see Section~\ref{subsec:LPRelax}), Q-tree search~\cite{Larsson2020} and k-class tree pruning~\cite{Kraetzschmar2004}.
The results are discussed in detail and brought into context of the topics discussed throughout this chapter.
We employ the Gurobi CVX solver~\cite{cvx,cvx2} to solve the information-theoretic integer program formulated in Section~\ref{sec:ITabsLP}.

\begin{figure}[h!]
	\centering
	\subfloat[]{\includegraphics[width=0.23\textwidth]{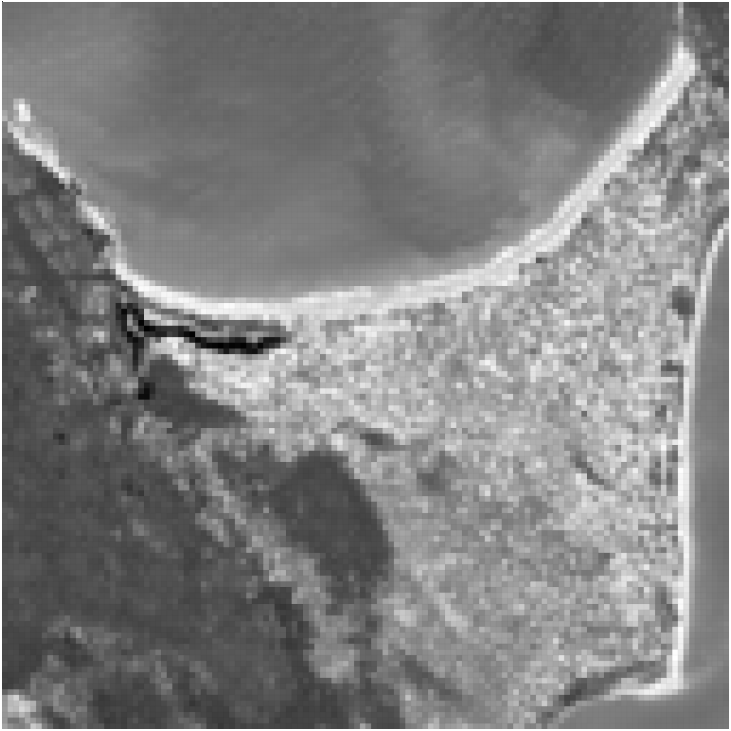}\label{fig:ex1_original}} \hspace{6pt}
	\subfloat[]{\includegraphics[width=0.23\textwidth]{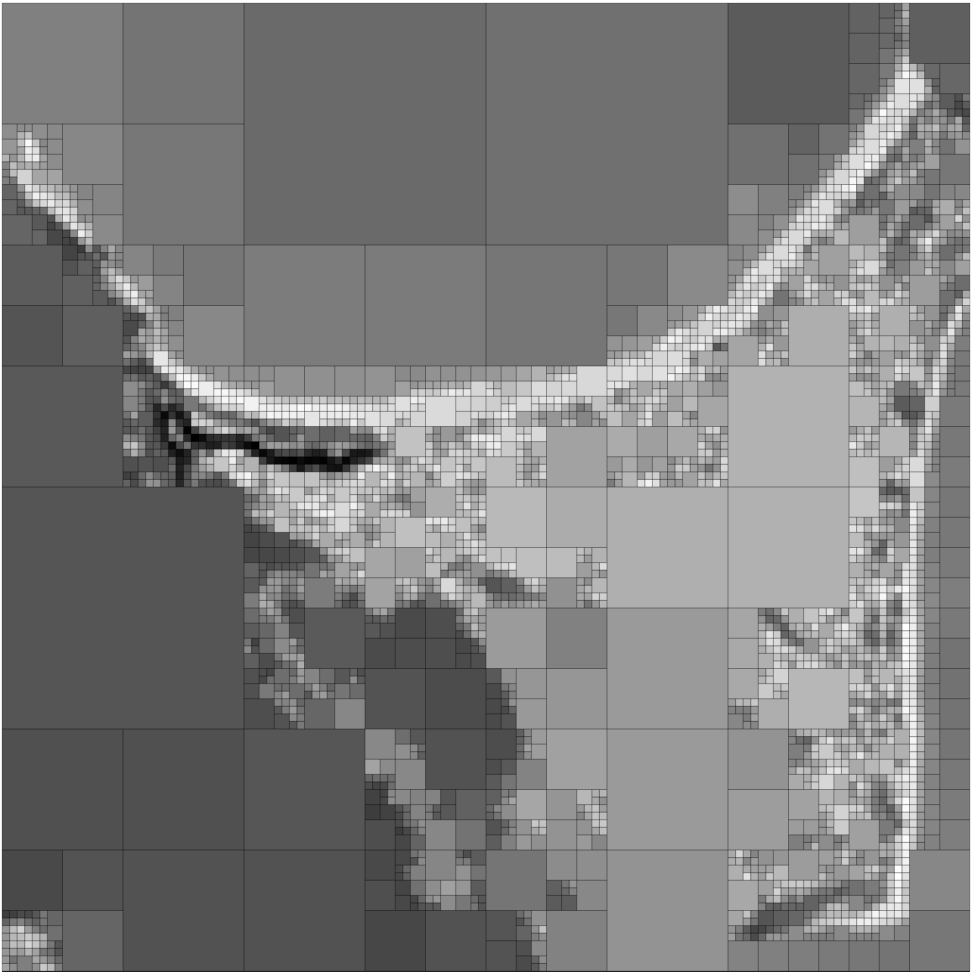}\label{fig:ex1_abs1}} \hspace{6pt}
	\subfloat[]{\includegraphics[width=0.23\textwidth]{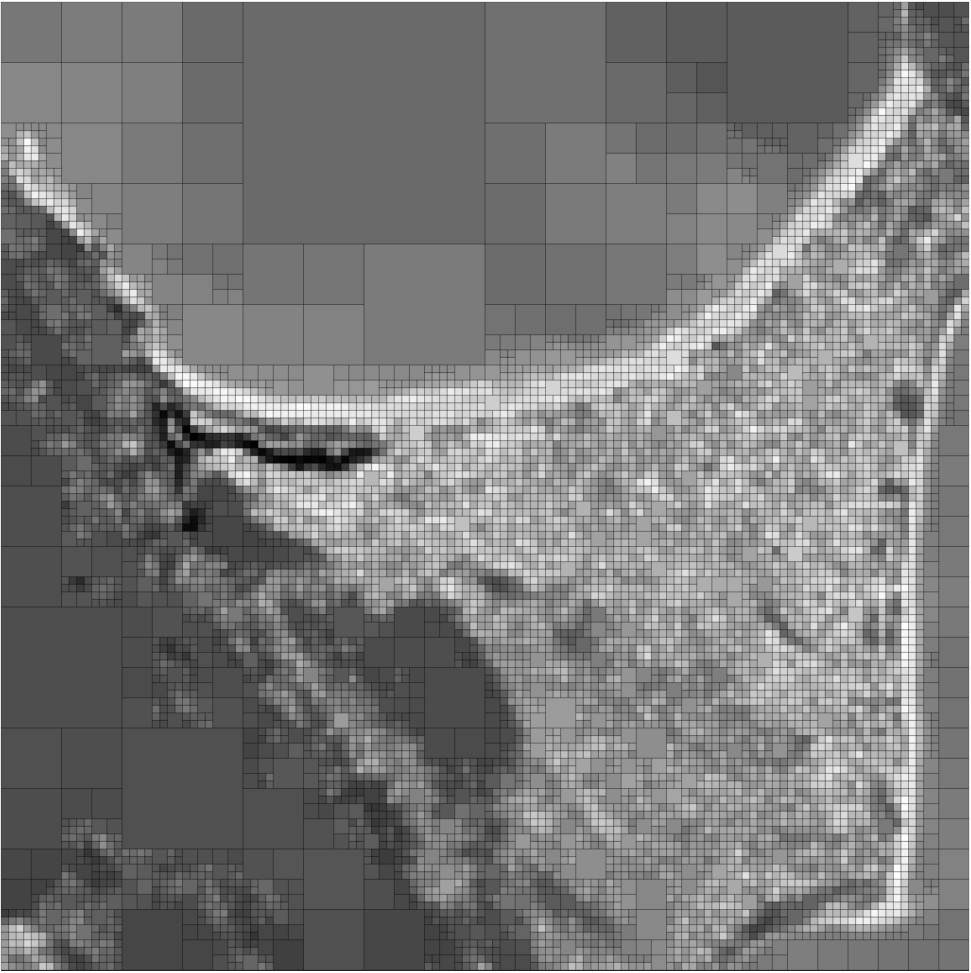}\label{fig:ex1_abs2}} \hspace{6pt}
	\subfloat[]{\includegraphics[width=0.23\textwidth]{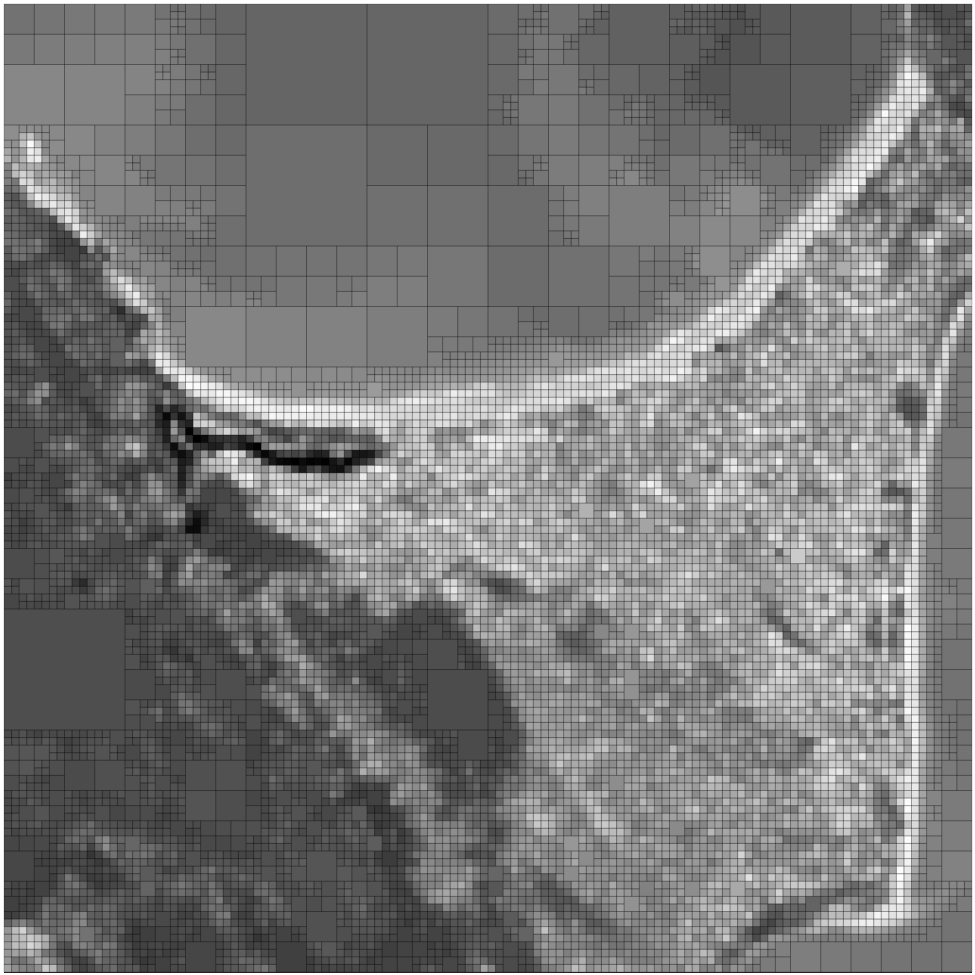}\label{fig:ex1_abs3}} 
	\caption{(a) Original, full resolution, \(128 \times 128\) image of the environment \(\W\), (b)-(d) abstractions of \(\W\) obtained by solving the integer program for a range of \(\hat D\) values. (b) Abstraction for \(\hat D = 0.0555\) nats (\(\nicefrac{\hat D}{I(X;Y)} = 0.965\)), which retains \(\nicefrac{I_Y(\T)}{I(X;Y)} = 0.9650\) of the available relevant information and contains \(34.9\%\) of the nodes compared to the full resolution image, (c) Abstraction for \(\hat D = 0.0574\) nats (\(\nicefrac{\hat D}{I(X;Y)} = 0.998\)), which retains \(\nicefrac{I_Y(\T)}{I(X;Y)} = 0.9984\) of the available relevant information and contains \(67.6\%\) of the nodes compared to the full resolution image, (d) abstraction for \(\hat D = 0.0575\) nats (\(\nicefrac{\hat D}{I(X;Y)} = 0.999\)), which retains \(\nicefrac{\hat D}{I(X;Y)} = 0.999\) of the available relevant information and contains \(76.9\%\) of the nodes as compared to the full resolution image. Notice that, in this example, with only approximately \(35\%\) of the nodes we obtain an abstraction that retains \(96.5\%\) of the relevant information.}
	\label{fig:ex1Abstractions}
\end{figure}

We first consider the satellite image shown in Fig.~\ref{fig:ex1_original}, which is of dimension \(128 \times 128\) (\(\ell = 7\)).
In this example, the relevant variable \(Y: \Omega \to \{0,1\}\) is assumed to be the color intensity of cells, where the outcome \(Y = 0\) corresponds to low intensity (lighter color) and the outcome \(Y = 1\) corresponds to high color intensity (darker color).
In addition, we let the uncompressed (original signal) variable be \(X: \Omega \to \{1,\ldots,\lvert \Nlf(\T_\W)\rvert\}\), whose outcomes \(x\) are the individual grid cells of the original image.
We assume the map represents the conditional distribution \(p(y=1|x)\), where each \(x\) is an outcome (grid cell) of \(X\), and \(p(y=1|x) \in [0,1]\) is the color intensity of the cell \(x \in \Nlf(\T_\W)\).
In general, the color intensity encoded by the map may be representative of a number of important aspects of the world, such as elevation data or probabilistic occupancy information.\footnote{In this case, \(p(y=1|x)\) represents the probability that the cell \(x \in \Nlf(\T_\W)\) is occupied. See~\cite{Larsson2020,Larsson2020b,Thrun2006} for more details on occupancy grids and abstraction.}
In both of these cases, the agent may not be able to determine, with certainty, the elevation data or occupancy information of grid cells due to limited on-board computational resources in processing camera images in order to create the environment map~\cite{Wang2019}.
Furthermore, agents may need to abstract, or compress, environment maps due a number of factors, including limited availability of on-board memory resources to store environment representations, or to transmit environment maps to other agents (e.g., satellite ground stations) across capacity-limited communication channels.

The input distribution \(p(x,y)\) is then created according to \(p(x,y) = p(y|x)p(x)\), where we assume throughout this chapter that the distribution \(p(x)\) is uniform (i.e., \(p(x) = \nicefrac{1}{\lvert \Nlf(\T_\W)\rvert}\)).
By choosing the distribution \(p(x)\) to be uniform, we perform what we call \textit{region agnostic} abstraction, whereby our framework will not favor any region of the map over another when generating the multi-resolution trees.
However, while we assume \(p(x)\) is uniform in this chapter for the purposes of demonstrating the utility of our approach, \textit{any} valid distribution \(p(x)\) may be used in our framework.
The use of a non-uniform \(p(x)\) will lead to \textit{region-specific} abstraction, as areas of the map with no support through \(p(x)\) do not contain any relevant information, irrespective of the distribution \(p(y|x)\).
Consequently, employing a non-uniform \(p(x)\) will result in refinement only in the regions for which \(p(x) > 0\).
For more information regarding region-specific abstraction in the context of our framework, we refer the interested reader to~\cite{Larsson2020}.

By solving the integer program~\eqref{eq:intLinProg2}-\eqref{eq:intLinProg2_Cons3} for a range of \(\hat D \geq 0\) values, we obtain the abstractions shown in Figs.~\ref{fig:ex1_abs1}-\ref{fig:ex1_abs3}.
From Figs.~\ref{fig:ex1_abs1}-\ref{fig:ex1_abs3} we see a number of interesting observations.
First, notice that, as the value of \(\hat D \geq 0\) is increased, the resolution of the environment representation increases.
This is expected, since the terms driving compression and information retention are in direct conflict, and thus the emergence of more informative solutions comes at the expense of a greater amount of \(X\)-information (i.e., less compression).
Secondly, observe that, at lower values of \(\hat D\), our framework retains those features of the image that are the most salient, leaving less noteworthy, finer-scale, details of \(\W\) to emerge at greater values of \(\hat D\).
To be more specific, consider the abstractions shown in Fig.~\ref{fig:ex1_abs1} and Fig.~\ref{fig:ex1_abs3}, where the former of these images is obtained at a lower value of \(\hat D\) than the latter.
While the abstraction Fig.~\ref{fig:ex1_abs1} contains less relevant information than that in Fig.~\ref{fig:ex1_abs3}, the two representations both show regions of the image where there is a stark difference in color intensity with great detail.

To understand why this is happening, we examine relations~\eqref{eq:condDistYinfoTree} and~\eqref{eq:finalDeltaTreeY}.
The first of these equations, namely~\eqref{eq:condDistYinfoTree}, shows how the values of \(p(y|t)\) for any node \(t \in \Nint(\T_\W)\) is computed from knowledge of \(p(y|t')\) of its children \(t'\in \chd(t)\).
The second relation, that of~\eqref{eq:finalDeltaTreeY}, then displays how the distributions \(p(y|t)\) and \(p(y|t')\) for \(t \in \Nint(\T_{\W}),~t'\in\chd(t)\) are utilized to compute the incremental change in relevant information contributed by node \(t\), given by \(\Delta \hat I_Y(t)\).
From these relations, we see that two components are required to compute \(\Delta \hat I_Y(t)\) for any \(t\in \Nint(\T_\W)\); namely \(p(t)\) and \(\mathrm{JS}_{\Pi}(p(y|t'_1),\ldots,p(y|t'_4))\).
Observe that, from the properties of the JS-divergence, the more diverse the distributions \(p(y|t')\) for \(t' \in \chd(t)\) are from their parent distribution \(p(y|t)\), the greater the value of \(\mathrm{JS}_{\Pi}(p(y|t'_1),\ldots,p(y|t'_4))\).
Moreover, from the relation~\eqref{eq:finalDeltaTreeY} we see that nodes \(t \in \Nint(\T_\W)\) that have the greatest changes in incremental information, \(\Delta \hat I_Y(t)\), are those nodes \(t\) that have large values of \(p(t)\) \emph{and} a high divergence (diversity) of the distributions \(p(y|t')\) of their children \(t' \in \chd(t)\), as measured by the JS-divergence.
In the context of our example this means that merging cells that are diverse with respect to their grey-scale intensity will have large values of \(\mathrm{JS}_{\Pi}(p(y|t'_1),\ldots,p(y|t'_1))\), whereas aggregating cells that are equal in their grey-scale intensity (i.e., are the same shade of grey), that is, \(p(y|t'_1) = \ldots = p(y|t'_4)\), will result in \(\mathrm{JS}_{\Pi}(p(y|t'_1),\ldots,p(y|t'_4)) = 0\).
Intuitively, when \(p(y|t'_1) = \ldots = p(y|t'_4)\), one's ability to predict the value of the color intensity is equally as good from the aggregated node as if the cells were refined and the resolution were increased.
In contrast, when there is a greater color intensity difference between nodes that are aggregated (i.e., \(\mathrm{JS}_{\Pi}(p(y|t'_1),\ldots,p(y|t'_4))\) is relatively large) then information regarding the color intensity of the merged nodes is lost, and can be recovered by splitting the aggregate cell and by increasing the resolution in the region.    
It is for the aforementioned reasons that regions in which there is a large color intensity difference are represented with relatively high resolution at lower values of \(\hat D\), whereas those areas with homogenous color intensities remain merged, or aggregated, to larger super nodes even at greater values of \(\hat D\).

\begin{figure}[t]
	\centering
	\includegraphics[width=0.55\textwidth]{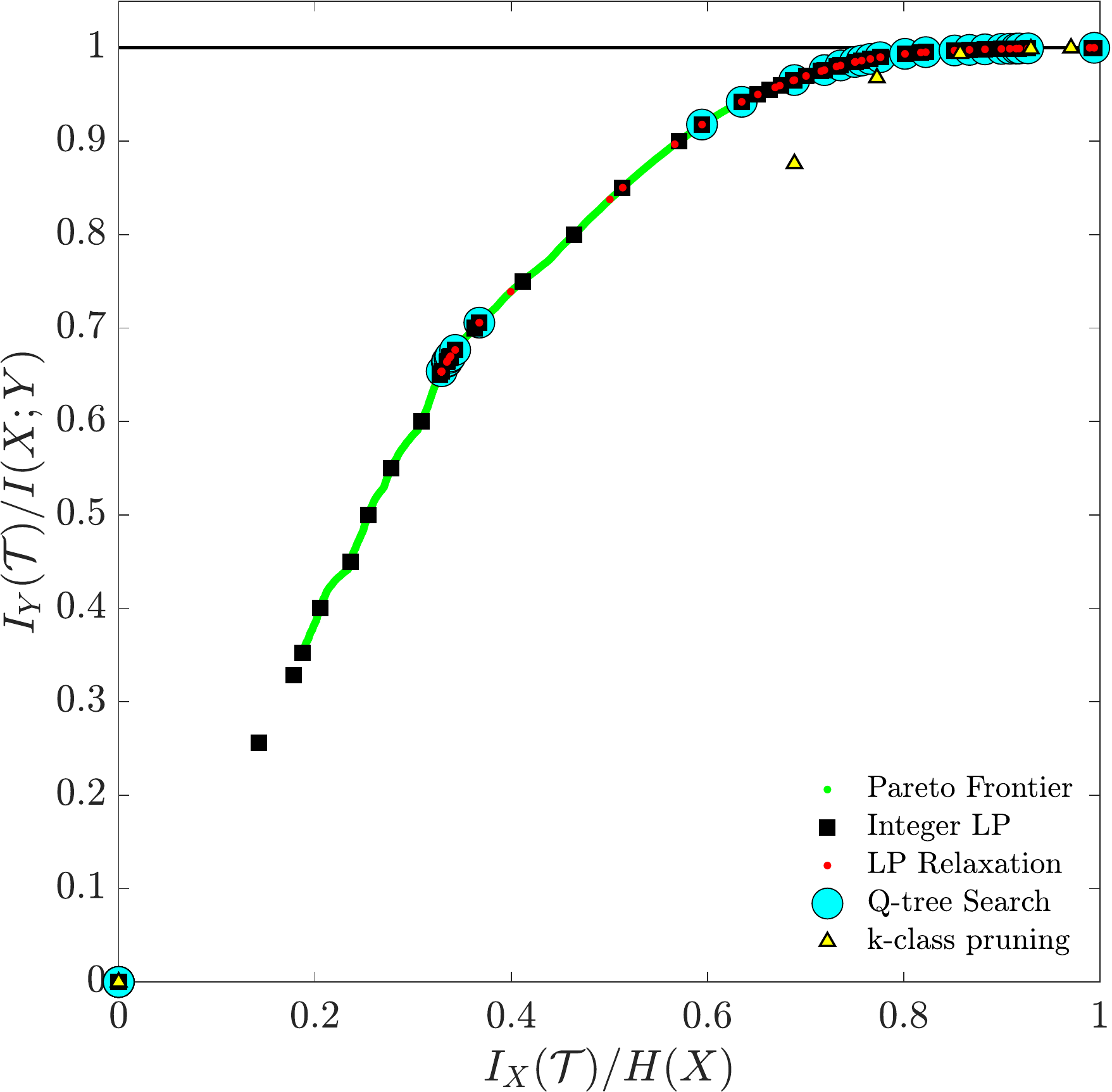}
	\caption{Normalized information plane for the environment shown in Fig.~\ref{fig:ex1_original}. The figure contains solutions (trees) obtained from the integer program, Q-tree search, LP relaxation and k-class tree pruning. In order to generate a valid multi-resolution tree representation of \(\W\) from the LP relaxation result, we apply the rule~\eqref{eq:extensionTruncationRule} with \(\delta = 0.5\).}
	\label{fig:ex1InformationPlane}
\end{figure}

From the results in Figs.~\ref{fig:ex1_abs1}-\ref{fig:ex1_abs3}, we also see that our framework finds multi-resolution trees that retain more than \(95\%\) of the relevant information and contain fewer than \(35\%\) of the nodes in the original representation.
Moreover, nearly all the relevant information can be retained by employing an abstraction that contains \(13\%\) fewer nodes than the full resolution image (see, Fig.~\ref{fig:ex1_abs3}).
These results represent a significant reduction in the number of nodes employed in the reduced representation, while sacrificing little in terms of relevant information retention.
To obtain a more holistic understanding of the trade-off between relevant information retention and compression for the environment in Fig.~\ref{fig:ex1_original}, we show in Fig.~\ref{fig:ex1InformationPlane} solutions obtained from the integer program~\eqref{eq:intLinProg2}-\eqref{eq:intLinProg2_Cons3} in the normalized information plane. 
In addition to the results obtained from executing the integer program, we also show in Fig.~\ref{fig:ex1InformationPlane} information plane points for other algorithms for abstraction, including Q-tree search, k-class trees and LP relaxation as well as the set of Pareto optimal values obtained by the method described in Section~\ref{subsec:ParetoEfficiency}.

From Fig.~\ref{fig:ex1InformationPlane} we observe that, in this example, the solutions returned by the LP relaxation, Q-tree search and the integer programming approaches\footnote{We note that the worst-case time complexity of the ILP and Q-tree search algorithms are as follows: ILP is NP-compelete~\cite[Ch.15]{Papadimitriou1998combinatorial} and Q-tree search is \(\mathcal O (\lvert \Nlf(\T_\W) \rvert)\)~\cite[pp.1680-1681]{Larsson2020}. Comparable performance of the two algorithms was observed for the examples presented in this chapter.} are all Pareto efficient.
The same cannot be said, however, regarding the k-class tree solutions, which lie below the collection of Pareto optimal values.
As a result, we conclude that the k-class tree pruning abstractions are sub-optimal, since the trees that are obtained from the k-class pruning approach retain less relevant information than trees returned by the integer programming developed in this chapter.
This is expected, as the k-class tree pruning method is an ad-hoc approach towards generating tree abstractions that is not driven by information-theoretic principles of information retention and compression.

Notice also that the solutions from the Q-tree search, integer programming and LP relaxation all show good agreement in the information plane.
As we increase the value of \(\hat D\), we move to the right in the information plane, obtaining solutions that retain more information regarding the relevant random variable, at the expense of achieving less compression in the resulting multi-resolution representation.
Thus, when \(\hat D = 0\), we obtain the trivial (root) tree \(\mathcal R_{\W}\) (which retains no relevant information, but maximizes compression), and as \(\hat D \to I(X;Y)\) trees that retain all the relevant information emerge as the solution to the integer program.
We recover a family of solutions (trees) for values of \(\hat D\) between these two extremes.
Similarly, the Q-tree search solutions move to the right in the information plane as the value of \(\beta > 0\) is increased, rendering trees that retain all the available relevant information as \(\beta \to \infty\). In the Q-tree search method, the trivial (root) abstraction is returned as \(\beta\) approaches \(0\).
Recall, however, that in contrast to the Q-tree search approach, the integer programming method developed in this chapter does not require the guesswork of tuning the \(\beta > 0\) parameter when designing abstractions to meet, for example, channel-capacity or memory resource constraints.
Instead, in the integer programming approach, these constraints can be directly incorporated into the problem formulation a priori, allowing for task-relevant abstractions to be obtained that are guaranteed to meet the imposed resource constraints.

\begin{figure}[t]
	\centering
	\subfloat[]{\includegraphics[width=0.23\textwidth]{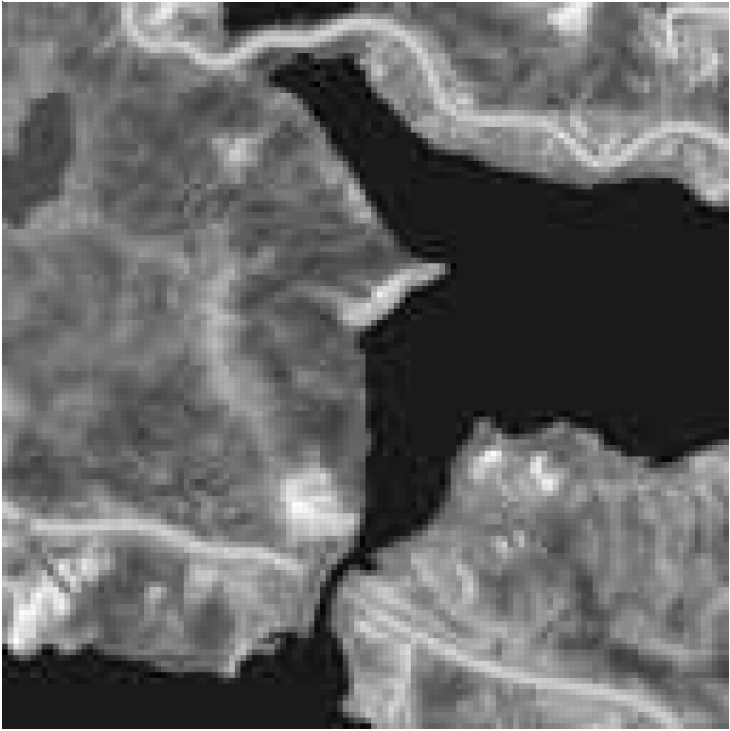}\label{fig:ex2_original}} \hspace{6pt}
	\subfloat[]{\includegraphics[width=0.23\textwidth]{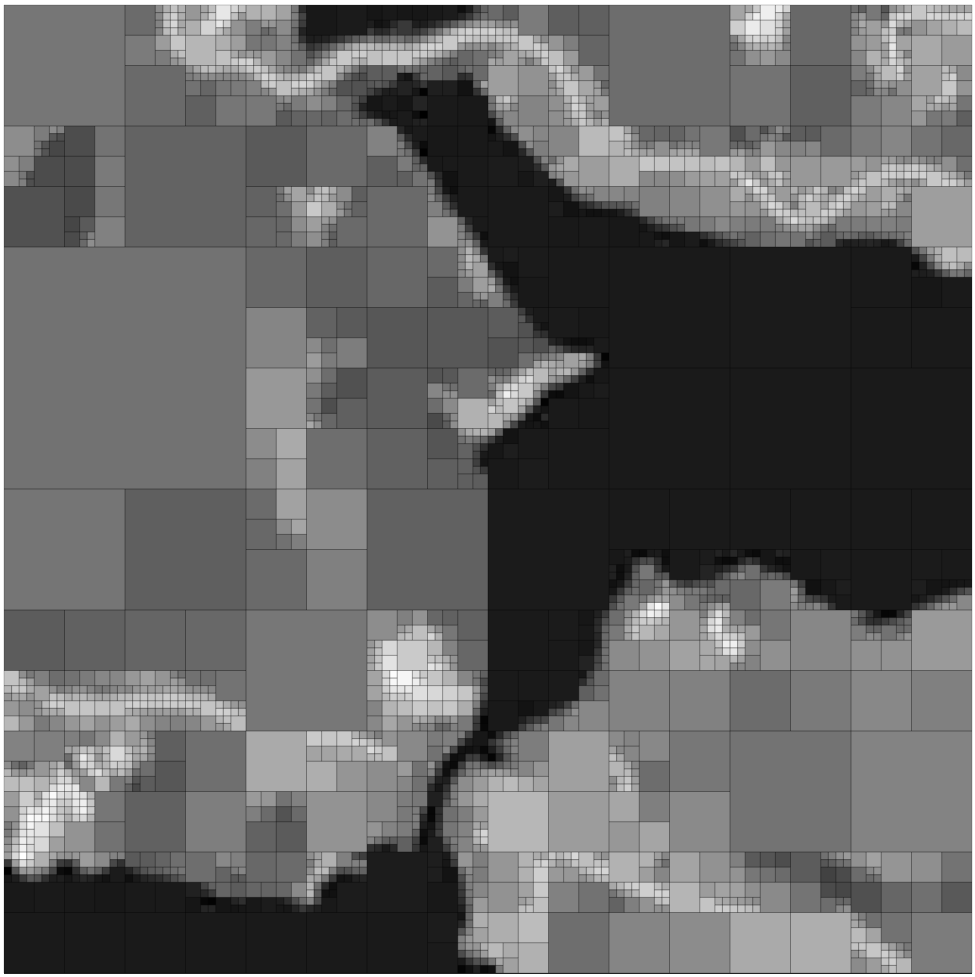}\label{fig:ex2_abs1}} \hspace{6pt}
	\subfloat[]{\includegraphics[width=0.23\textwidth]{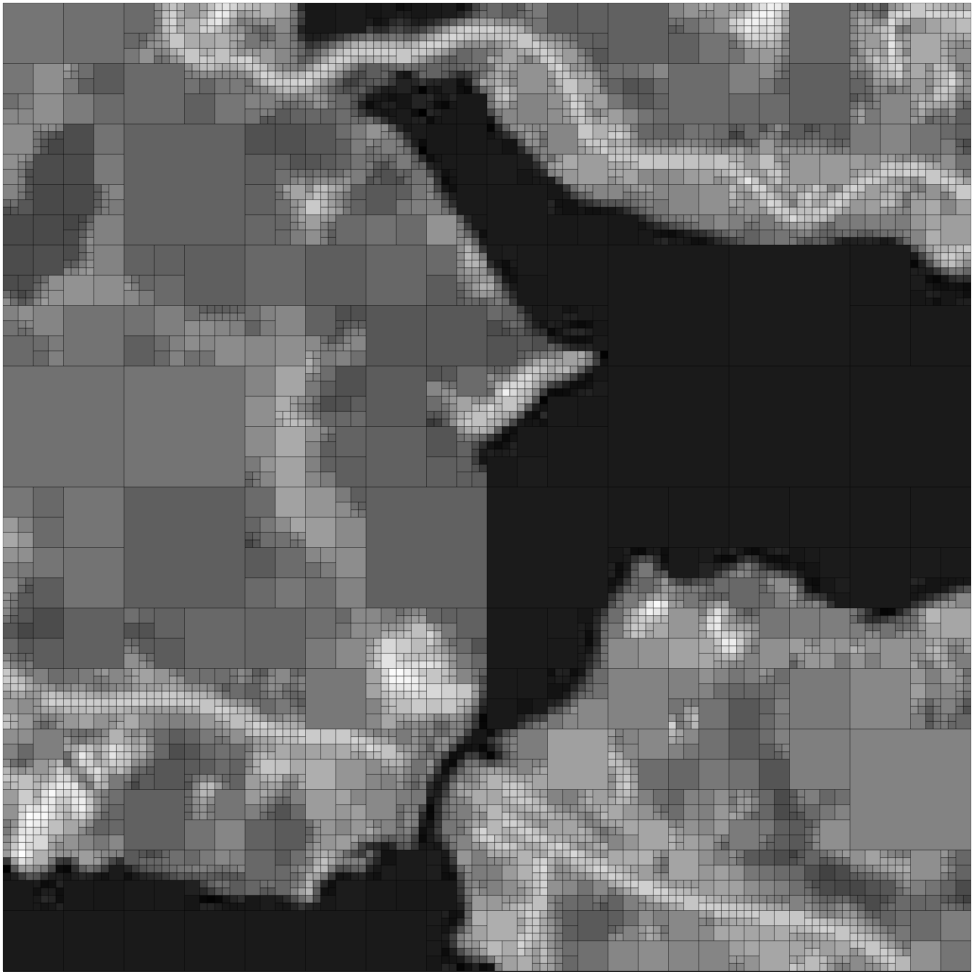}\label{fig:ex2_abs2}} \hspace{6pt}
	\subfloat[]{\includegraphics[width=0.23\textwidth]{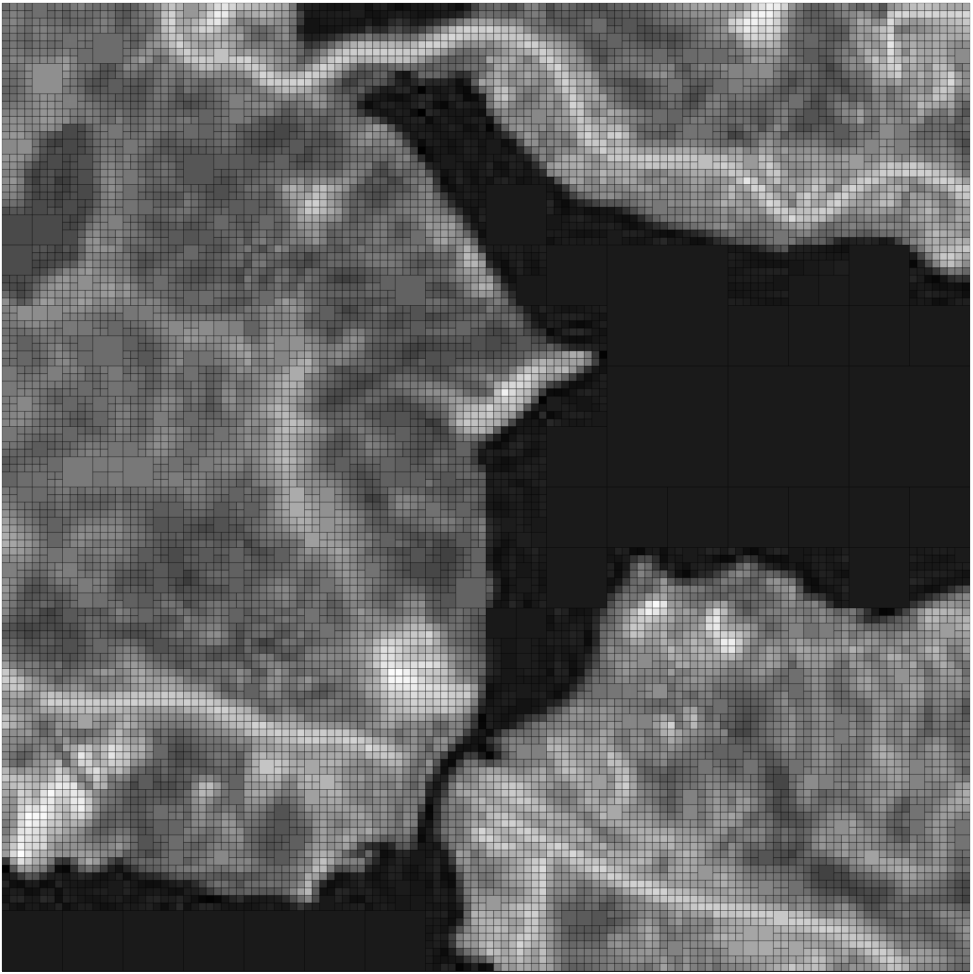}\label{fig:ex2_abs3}} 
	\caption{(a) Original, full resolution, \(128 \times 128\) image of the environment \(\W\), (b)-(d) abstractions of \(\W\) obtained by solving the integer program for a range of \(\hat D\) values. (b) Abstraction for \(\hat D = 0.0954\) nats (\(\nicefrac{\hat D}{I(X;Y)} = 0.959\)), which retains \(~\nicefrac{I_Y(\T)}{I(X;Y)} = 0.959\) of the available relevant information and contains \(20.2\%\) of the nodes as compared to the full resolution image, (c) Abstraction for \(\hat D = 0.0981\) nats (\(\nicefrac{\hat D}{I(X;Y)} = 0.985\)), which retains \(\nicefrac{I_Y(\T)}{I(X;Y)} = 0.985\) of the available relevant information and contains \(35.5\%\) of the nodes as compared to the full resolution image, (d) Abstraction for \(\hat D = 0.0995\) nats (\(\nicefrac{\hat D}{I(X;Y)} = 0.99\)), which retains \(\nicefrac{I_Y(\T)}{I(X;Y)} = 0.99\) of the available relevant information and contains \(79.4\%\) of the nodes as compared to the full resolution image.
	}
	\label{fig:ex2Abstractions}
\end{figure}

Lastly, we notice that there are points in the information plane where the solution to the integer program ``jumps" from one point to another.
For example, no integer programming solutions are found between the information plane points of the origin \((0,0)\) (i.e., the trivial abstraction) and \((0.1429, 0.2559)\).
The resulting ``kink'' is due to the discrete nature of the solution space.
Specifically, in this example, no solutions are obtained between the origin and \((0.1429, 0.2559)\) as, from the root tree/node \(\mathcal R_{\W}\), only one expansion is permitted -- namely, expanding the root itself.
Expanding the root results in a finite amount of information gain, captured by \(\Delta \hat I_X(t)\) and \(\Delta \hat I_Y(t)\), respectively, where \(t = \mathcal R_{\W}\).
Thus, for \(\hat D \in (0,\Delta\hat I_Y(t)]\) (where \(t = \mathcal R_{\W}\)), the resulting solution to the integer program will be the root-expanded tree, resulting in a jump in the information plane from \((0,0)\) to \((\Delta \hat I_X(t), \Delta \hat I_Y(t)) = (0.1429, 0.2559)\), as seen in Fig.~\ref{fig:ex1InformationPlane}.

We also apply our approach to the environment shown in Fig.~\ref{fig:ex2_original}, which is of dimension \(128 \times 128\) (\(\ell = 7\)).
In order to provide a useful discussion comparing and contrasting the results obtained by employing our framework to design abstractions for this new environment, we again assume \(p(x)\) is a uniform distribution and  \(p(x,y) = p(y|x)p(x)\), where the map encodes the conditional distribution \(p(y|x)\).
By solving the information-theoretic integer program for various values of \(\hat D\), we obtain the abstractions shown in Figs.~\ref{fig:ex2_abs1}-\ref{fig:ex2_abs3}.

From the results in Figs.~\ref{fig:ex2_abs1}-\ref{fig:ex2_abs3} we see a number of trends that are similar to what it was observed in the prior example.
To this end, we first notice that, as the value of \(\hat D\) is increased, the resolution of the resulting abstract representation increases.
Furthermore, from Fig.~\ref{fig:ex2Abstractions} we observe that employing an abstraction with only approximately \(20\%\) of the resolution of the original image, we are able to retain more than \(95\%\) of the relevant information.
Consequently, many of the cells from the original image may be merged with little degradation in ones ability to predict the color intensity of cells.
Moreover, we see that at lower values of \(\hat D\), our framework finds trees that maintain high resolution in areas where there is a large change in color intensity between finest-resolution cells.
For example, notice that the multi-resolution tree depicted in Fig.~\ref{fig:ex2_abs1} aggregates cells in areas where there is relatively small change in color intensity (e.g., the homogenously dark or lighter portions of the map) while regions along the boundaries of light to dark cells are shown in high resolution.
As the value of \(\hat D\) is increased, the priority of  multi-resolution abstractions becomes retaining as much information as possible, resulting in refinement of those regions where there is a color intensity difference, but that were aggregated at lower values of \(\hat D\).
Lastly, notice that the dark region of the map remains aggregated into super cells even at larger values of \(\hat D\).
Per our previous discussion regarding the incremental change in relevant information, we note that this occurs as areas of homogeneous color intensity do not contain any relevant information, and thus the dark portion of the map remains aggregated into as large cells as possible while ensuring the resulting multi-resolution tree is a valid quadtree representation of \(\W\).

\begin{figure}[t]
	\centering
	\includegraphics[width=0.55\textwidth]{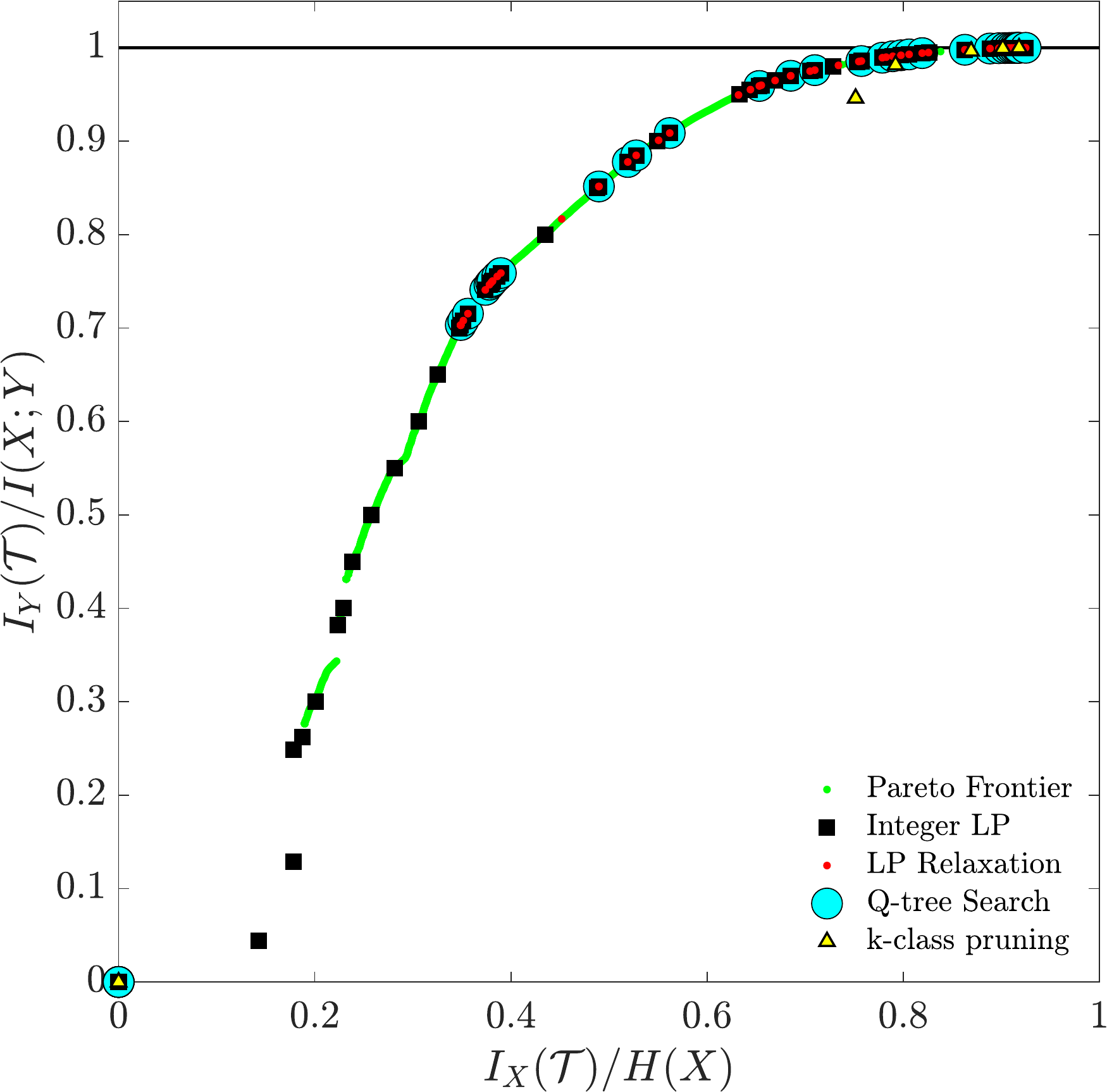}
	\caption{Normalized information plane for the environment shown in Fig.~\ref{fig:ex2_original}. The figure contains solutions (trees) obtained from the integer program, Q-tree search, LP relaxation and k-class tree pruning. In order to generate a valid multi-resolution tree representation of \(\W\) from the LP relaxation result, we apply the rule~\eqref{eq:extensionTruncationRule} with \(\delta = 0.5\).}
	\label{fig:ex2InformationPlane}
\end{figure}

Shown in Fig.~\ref{fig:ex2InformationPlane} are the information plane points obtained from applying other aggregation algorithms, such as Q-tree search and k-class tree pruning, to the example environment in Fig.~\ref{fig:ex2_original}.
We again observe good agreement between the solutions obtained from the Q-tree search, LP relaxation and integer programming approaches.
Interestingly, we see from the integer program solutions shown in Fig.~\ref{fig:ex2InformationPlane} that there is one identifiable solution that is not Pareto efficient.
Notice that the third and fourth integer solutions both have the same value of \(I_X(\T)\) (that is, they are along the same vertical line in the information plane), however, the fourth solution contains more relevant information than the third.
As a result, the third solution is not Pareto efficient (see Section~\ref{subsec:ParetoEfficiency}), since there is another feasible tree that, for the same amount of compression, retains more relevant information.
We note that this is the only one of the solutions obtained and displayed in Fig.~\ref{fig:ex2InformationPlane} that are not Pareto efficient, apart from those obtained via the k-class tree pruning method.
Furthermore, we once again observe that the k-class approach results in trees that are sub-optimal in their balance between compression and information retention.
Moreover, as was the case in Fig.~\ref{fig:ex1InformationPlane}, we notice that the solutions returned by the k-class tree method are grouped at the extreme ends of the information plane.
That is, the k-class solutions jump from the origin (which is the trivial compression obtained when only a single class is employed) to the upper right region of the information plane, which is the area where high-cardinality, highly informative solutions lie.
Consequently, while the k-class approach allows one to reduce memory storage requirements by creating abstractions of the environment, it does not allow for a diverse family of solutions to be obtained that gradually trade compression and information retention.
In contrast, our approach allows for a more principled trade-off between information retention and compression.

\section{Conclusion}

In this chapter, we considered the problem of designing task-relevant abstractions of operating environments for autonomous systems. 
The abstractions emerge as a function of agent resource constraints and are in the form of multi-resolution, hierarchical, tree structures that are not provided a priori.
To design the abstractions, we employ an information-theoretic approach that leverages the information bottleneck (IB) principle to formulate an optimization problem that searches for encodings (abstractions) of the environment that are maximally retentive regarding task-relevant information while maximizing compression.
In order to rigorously formulate our problem, we elucidate the connections between signal encoders and multi-resolution tree structures and detail how a tree abstraction problem can be viewed as an optimal encoder search, akin to information-theoretic approaches of signal compression.

Moreover, we show how designing information-theoretically driven tree abstractions that are maximally retentive regarding task-relevant information can be realized as an integer linear program by exploiting the structure of the problem.
The properties of the resulting integer program are discussed in detail, and a method for linear programming relaxation is presented, thereby allowing information-theoretic multi-resolution tree abstractions to be obtained by solving a (convex) linear optimization problem.
Numerical examples are presented that show the utility of the approach.
Furthermore, we compare and contrast our approach to existing methods for abstraction, including Q-tree search~\cite{Larsson2020} and k-class tree pruning~\cite{Kraetzschmar2004}.
The numerical results are discussed in detail and show the usefulness of our approach for generating task-relevant abstractions for resource-constrained autonomous systems.

\bibliographystyle{ieeetr}

\begin{thebibliography}{10}

\bibitem{Ponsen2010}
M.~Ponsen, M.~E. Taylor, and K.~Tuyls, ``Abstraction and generalization in
	reinforcement learning: A summary and framework,'' in {\em Adaptive and
	Learning Agents}, pp.~1--32, Springer Berlin Heidelberg, 2010.

\bibitem{Zucker2003}
J.~Zucker, ``A grounded theory of abstraction in artificial intelligence,''
	{\em Philosophical Transactions of the Royal Society of London, Series B:
	Biological Sciences 358}, no.~1435, pp.~1293--1309, 2003.

\bibitem{Holte2003}
R.~C. Holte and B.~Y. Choueiry, ``Abstraction and reformulation in artificial
	intelligence,'' {\em Philosophical Transactions of the Royal Society of
	London B: Biological Sciences}, vol.~358, pp.~1197--1204, July 2003.

\bibitem{Larsson2020b}
D.~T. Larsson, D.~Maity, and P.~Tsiotras, ``Information-theoretic abstractions
	for planning in agents with computational constraints,'' {\em IEEE Robotics
	and Automation Letters}, vol.~6, no.~4, pp.~7651--7658, 2021.

\bibitem{Cover2006}
T.~M. Cover and J.~A. Thomas, {\em Elements of Information Theory}.
\newblock John Wiley \& Sons, 2nd~ed., 2006.

\bibitem{Tishby1999}
N.~Tishby, F.~C. Pereira, and W.~Bialek, ``The information bottleneck method,''
	in {\em Allerton Conference on Communication, Control and Computing},
	(Monticello, IL, USA), pp.~368--377, 1999.
\newblock September 22-24, 1999.

\bibitem{GiladBachrach2003}
R.~Gilad-Bachrach, A.~Navot, and N.~Tishby, ``An information theoretic tradeoff
	between complexity and accuracy,'' in {\em Learning Theory and Kernel
	Machines}, pp.~595--609, Springer Berlin Heidelberg, 2003.

\bibitem{Strouse2017}
D.~Strouse and D.~J. Schwab, ``The deterministic information bottleneck,'' {\em
	Neural Computation}, vol.~29, pp.~1611--1630, jun 2017.

\bibitem{Chechik2002}
G.~Chechik and N.~Tishby, ``Extracting relevant structures with side
	information,'' in {\em Conference on Neural Information Processing Systems},
	(Vancouver, BC, CA), pp.~881--888, Dec. 2002.
\newblock December 9-14, 2002.

\bibitem{Tsiotras2007}
P.~Tsiotras and E.~Bakolas, ``A hierarchical on-line path planning scheme using
	wavelets,'' in {\em European Control Conference}, (Kos, GR), pp.~2806--2812,
	2007.
\newblock July 2-5, 2007.

\bibitem{Tsiotras2011}
P.~Tsiotras, D.~Jung, and E.~Bakolas, ``Multiresolution hierarchical
	path-planning for small {UAVs} using wavelet decompositions,'' {\em Journal
	of Intelligent {\&} Robotic Systems}, vol.~66, pp.~505--522, sep 2011.

\bibitem{Hauer2019}
F.~Hauer, {\em Path-{P}lanning {A}lgorithms in {H}igh-{D}imensional {S}paces}.
\newblock PhD thesis, Georgia Institute of Technology, 2019.

\bibitem{Hauer2015}
F.~Hauer, A.~Kundu, J.~M. Rehg, and P.~Tsiotras, ``Multi-scale perception and
	path planning on probabilistic obstacle maps,'' in {\em IEEE International
	Conference on Robotics and Automation}, (Seattle, WA, USA), pp.~4210--4215,
	2015.
\newblock May 26-30, 2015.

\bibitem{Cowlagi2012}
R.~V. Cowlagi and P.~Tsiotras, ``Multiresolution motion planning for autonomous
	agents via wavelet-based cell decompositions,'' {\em {IEEE} Transactions on
	Systems, Man, and Cybernetics, Part B (Cybernetics)}, vol.~42,
	pp.~1455--1469, oct 2012.

\bibitem{Cowlagi2011}
R.~V. Cowlagi, {\em Hierarchical Motion Planning for Autonomous Aerial and
	Terrestrial Vehicles}.
\newblock PhD thesis, Georgia Institute of Technology, 2011.

\bibitem{Cowlagi2010}
R.~V. Cowlagi and P.~Tsiotras, ``Multi-resolution path planning: Theoretical
	analysis, efficient implementation, and extensions to dynamic environments,''
	in {\em IEEE Conference on Decision and Control}, (Atlanta, GA, USA),
	pp.~1384--1390, 2010.
\newblock December 15-17, 2010.

\bibitem{Cowlagi2008}
R.~V. Cowlagi and P.~Tsiotras, ``Multiresolution path planning with wavelets: A
	local replanning approach,'' in {\em American Control Conference}, (Seattle,
	WA, USA), pp.~1220--1225, jun 2008.
\newblock June 11-13, 2008.

\bibitem{Slonim2002}
N.~Slonim, {\em The Information Bottleneck: Theory and Applications}.
\newblock PhD thesis, The Hebrew University, 2002.

\bibitem{Bondy1976}
J.~A. Bondy and U.~S.~R. Murty, {\em Graph Theory with Applications}.
\newblock Macmillan Education {UK}, 1976.

\bibitem{Hornung2013}
A.~Hornung, K.~M. Wurm, M.~Bennewitz, C.~Stachniss, and W.~Burgard, ``Ocotomap:
	An efficient probabilistic 3d mapping framework based on octrees,'' {\em
	Autonomous Robots}, vol.~34, pp.~189--206, Apr. 2013.

\bibitem{Einhorn2011}
E.~Einhorn, C.~Schr\"{o}ter, and H.-M. Gross, ``Finding the adequate resolution
	for grid mapping - cell sizes locally adapting on-the-fly,'' in {\em IEEE
	Conference on Robotics and Automation}, (Shanghai, CN), pp.~1843--1848, 2011.
\newblock May 9-13, 2011.

\bibitem{Kambhampati1986}
S.~Kambhampati and L.~S. Davis, ``Multiresolution path planning for mobile
	robots,'' {\em IEEE Journal of Robotics and Automation}, vol.~RA-2,
	pp.~135--145, September 1986.

\bibitem{Hauer2016}
F.~Hauer and P.~Tsiotras, ``Reduced complexity multi-scale path-planning on
	probabilistic maps,'' in {\em {IEEE} International Conference on Robotics and
	Automation}, (Stockholm, SE), pp.~83--88, may 2016.
\newblock May 16-21, 2016.

\bibitem{Kraetzschmar2004}
G.~K. Kraetzschmar, G.~P. Gassull, and K.~Uhl, ``Probabilistic quadtrees for
	variable-resolution mapping of large environments,'' {\em IFAC Proceedings
	Volumes}, vol.~37, pp.~675--680, July 2004.

\bibitem{Larsson2020}
D.~T. Larsson, D.~Maity, and P.~Tsiotras, ``Q-tree search: An
	information-theoretic approach toward hierarchical abstractions for agents
	with computational limitations,'' {\em {IEEE} Transactions on Robotics},
	vol.~36, no.~6, pp.~1669--1685, 2020.

\bibitem{Slonim2000}
N.~Slonim and N.~Tishby, ``Agglomerative information bottleneck,'' in {\em
	Conference on Neural Information Processing Systems}, (Denver, CO, USA),
	pp.~617--623, 1999.
\newblock November 29 - December 4, 1999.

\bibitem{larsson2021information}
D.~T. Larsson, D.~Maity, and P.~Tsiotras, ``Information-theoretic abstractions
	for resource-constrained agents via mixed-integer linear programming,'' in
	{\em Proceedings of the Workshop on Computation-Aware Algorithmic Design for
	Cyber-Physical Systems}, (Nashville, TN, USA), pp.~1--6, 2021.
\newblock May 19-21, 2021.

\bibitem{Bertsekas2005}
D.~Bertsekas, {\em Dynamic programming and optimal control}.
\newblock Athena Scientific, 2005.

\bibitem{SuttonBarto}
R.~S. Sutton and A.~G. Barto, {\em Reinforcement Learning: An Introduction}.
\newblock MIT Press, 1998.

\bibitem{Lin1991}
J.~Lin, ``Divergence measures based on the {S}hannon entropy,'' {\em IEEE
	Transactions on Information Theory}, vol.~37, pp.~145--151, January 1991.

\bibitem{Boyd2004convex}
S.~Boyd and L.~Vandenberghe, {\em Convex Optimization}.
\newblock Cambridge university press, 2004.

\bibitem{Das1997}
I.~Das and J.~E. Dennis, ``A closer look at drawbacks of minimizing weighted
	sums of objectives for pareto set generation in multicriteria optimization
	problems,'' {\em Structural Optimization}, vol.~14, no.~1, pp.~63--69, 1997.

\bibitem{Papadimitriou1998combinatorial}
C.~H. Papadimitriou and K.~Steiglitz, {\em Combinatorial Optimization:
	Algorithms and Complexity}.
\newblock Courier Corporation, 1998.

\bibitem{Ahuja1988network}
R.~K. Ahuja, T.~L. Magnanti, and J.~B. Orlin, {\em Network Flows}.
\newblock Pearson, 1993.

\bibitem{cvx}
M.~Grant and S.~Boyd, ``{CVX}: Matlab software for disciplined convex
	programming, version 2.1.'' \url{http://cvxr.com/cvx}, Mar. 2014.

\bibitem{cvx2}
M.~Grant and S.~Boyd, ``Graph implementations for nonsmooth convex programs,''
	in {\em Recent Advances in Learning and Control} (V.~Blondel, S.~Boyd, and
	H.~Kimura, eds.), Lecture Notes in Control and Information Sciences,
	pp.~95--110, Springer-Verlag Limited, 2008.

\bibitem{Thrun2006}
S.~Thrun, W.~Burgard, and D.~Fox, {\em Probabilistic Robotics}.
\newblock MIT Press, 2006.

\bibitem{Wang2019}
Y.~Wang, Z.~Lai, G.~Huang, B.~H. Wang, L.~van~der Maaten, M.~Campbell, and
	K.~Q. Weinberger, ``Anytime stereo image depth estimation on mobile
	devices,'' in {\em International Conference on Robotics and Automation},
	(Montreal, QC, CA), pp.~5893--5900, 2019.
\newblock May 20-24, 2019.
\end{thebibliography}


\end{document}